\documentstyle[preprint,tighten,eqsecnum,floats,epsfig,aps]{revtex}

\def\beq{\begin{eqnarray}}
\def\eeq{\end{eqnarray}}
\def\ks{\rlap/k}
\def\MD{m_\Delta}

\def\meno{{\LARGE \bf (-)}}

\begin{document}

\title{Contrasting $N \rightarrow \Delta$ and $N \rightarrow N$ \\ 
parity-violating asymmetries in nuclei}

\author{P. Amore$^\dagger$, R. Cenni$^\clubsuit$, T.W. Donnelly$^{\natural}$,
A. Molinari$^\spadesuit$}
\address{$^\dagger${\sl College of William and Mary, \\
Williamsburg, VA 23185, USA} \\
$^\clubsuit${\sl Dipartimento di Fisica, \\
Universit\'a di Genova and INFN, sezione
di Genova, Genova, Italy} \\
$^\natural${\sl Center for Theoretical Physics, \\
Laboratory for Nuclear Science and Department of Physics \\
Massachusetts Institute of Technology,\\
Cambridge, MA 02139-4307, USA} \\
$^\spadesuit${\sl Dipartimento di Fisica Teorica, \\
Universit\'a di Torino and INFN, sezione
di Torino, Torino, Italy}}

\maketitle

\begin{abstract}
Several perspectives that can be opened by studying
the electroexcitation of the delta via parity-violating 
electron scattering from nuclei are examined, working 
within the context of the relativistic Fermi gas model. 
A strong enhancement of the left-right asymmetry in the 
delta sector compared with that in the quasi-elastic 
regime is found and the potential to find clear 
signatures for the axial-vector contributions of the 
nucleon-to-delta transition to the asymmetry identified 
at specific low momentum transfer kinematics. Possibilities 
of probing the deformation of the delta are also explored, 
and using both the proton and nuclei as targets, the 
abilitiy to study the asymmetry on neutrons is studied.

\end{abstract}

\vspace{1cm}






\section{Introduction}

In this paper we study the asymmetry ${\cal A}$ measured in 
the parity-violating 
nuclear scattering of longitudinally polarized electrons in the region of 
excitation where the $\Delta (1232)$ dominates.
This theme has already been explored in the case of the single 
proton\cite{Muk98}; here we focus on scattering from nuclei and work
within the context of the relativistic Fermi gas (RFG) model.
Moreover, since the features that characterize 
the asymmetry in the region of the
$\Delta$ peak emerge more transparently when comparisons are made 
with the same observable at the quasi-elastic peak (QEP), 
we shall explore both domains in parallel. 
In particular, we shall focus on the energy behavior of $\cal A$ as one 
goes from the QEP domain to the $N \rightarrow \Delta$ region for a few 
values both of the momentum $q$ transferred by the electron to the nucleus
and of the electron scattering angle $\theta$. We shall see that a large 
increase of the many-body content in ${\cal A}$ occurs in the $\Delta$ 
sector as compared with the QEP region {\sl for small values of $\theta$}.

A study of $\cal A$ requires knowledge of the response functions, 
both parity-conserving (PC, electromagnetic) and parity-violating 
(PV, weak neutral current), and accordingly we 
proceed within the context of the RFG model via the
polarization propagator method employing the $\gamma N \Delta$ vertices of 
Devenish et al. \cite{Dev76}. 
We thus recover the PC $N \rightarrow\Delta$ longitudinal and transverse 
responses obtained previously in \cite{Ama99},  and in addition we 
obtain the axial $N \rightarrow \Delta$ response. It turns out that the 
asymmetry offers some hope of disentangling --- at least in the $\Delta$ 
region --- the otherwise quite elusive nuclear axial response.

The results referred to above are obtained initially by viewing the 
$\Delta$ as a stable particle. It is worth pointing out that in this 
scheme when the $\Delta$ mass 
$m_\Delta$ approaches the nucleon mass $m_N$ the 
$N \rightarrow \Delta$ RFG responses evolve into the corresponding 
$N \rightarrow N$ quasi-elastic RFG responses except, 
of course, for appropriate 
changes in the form factors. These formal relationships between 
the two sets of responses are of relevance in connection with 
nuclear $y$-scaling, as we shall illustrate later.

To explore the impact of the finite lifetime of the 
$\Delta$ on our findings in a few instances
we then compute the asymmetry, ascribing to the 
$\Delta$ a width $\Gamma$ and folding the RFG responses obtained at
zero width with a phenomenological $\Delta$ distribution. While our results 
for the asymmetry turn out be relatively unaffected by $\Gamma$, it is also 
obviously apparent that an appropriate width is required to account 
satisfactorily for the actual response functions.

A further theme addressed in the present work relates to the longitudinal 
PC $N \rightarrow \Delta$ response: this is contributed to by a presently 
poorly known Coulomb operator, as well as by the magnetic operator driven 
by relativity and Fermi motion \cite{Chan93}. To assess the
importance of the latter versus the former in the RFG framework we
proceed as previously for its counterpart in the QEP sector and
reduce the longitudinal $N \rightarrow \Delta$ response, so deriving 
a kind of transverse Coulomb 
sum rule in the $\Delta$ domain which saturates at large momentum 
tranfer ($\approx \ 2.5 \ {\rm GeV/c}$) to a value set by the Fermi momentum
$k_F$.

In connection with the asymmetry, such PV studies may provide new insight
into specific aspects of nuclear dynamics. Accordingly we assess the
feasibility of actually measuring the asymmetry in the $N \rightarrow \Delta$ 
domain by estimating the attainable precision when detecting $\cal A$ at 
kinematics relevant for TJNAF. Notably this precision  turns out to be
close to, if not better than, the one that can be reached 
in the region of the QEP \cite{Don92}.

Given the feasibility of such measurements the question arises: Is there
an advantage to exploring the $N \rightarrow \Delta$ asymmetry in complex 
nuclei rather than only in the proton? 
We search for an answer to this question by working out the RFG predictions 
in the limit of vanishing $k_F$. We find that in the $\Delta$ sector the 
RFG asymmetry displays a minor $k_F$ dependence, as far as  its magnitude 
is concerned, remaining essentially equal to the single-particle asymmetry 
over a wide range of $k_F$, no matter what the momentum transfer 
or scattering angle $\theta$ are. However, this is not seen to be the case
in the QEP sector where the Fermi motion and the isoscalar-isovector 
competition conspire to yield in the asymmetry an interesting and 
channel-dependent $k_F$ dependence.

The present study, being confined to the RFG model, does not take into
account $N N$ or $N \Delta$ correlations and MEC contributions 
for the various responses.
It should also be added that we have neglected the non-resonant (background) 
pionic contribution to the physics in the $\Delta$ sector, whose relevance 
should still be assessed. 
Finally, in this work we had to reckon with the poor knowledge presently 
available on most of the $N \rightarrow \Delta$ form factors. {\sl Faut de mieux} 
for the axial sector we have relied on the one hand on the Adler scaling 
hypothesis \cite{Adl68} for the electric form factor and, on the other, 
on the constituent quark model, which appears to predict  an axial magnetic 
form factor that is substantially smaller than the electric one \cite{Hem95}. 
In the vector sector the electric and Coulomb form factors are poorly known: 
here we adopt the parametrization recently suggested in \cite{Ama99} to 
gain a first orientation on their role in the $N \rightarrow \Delta$ responses.

\section{The $\Delta$-hole polarization propagator in the RFG}

In this section  we pave the way to the calculation of the $N \rightarrow \Delta$ 
response functions within the framework of the symmetric RFG. 
These have been already computed in the vector sector in \cite{Ama99}; 
here we provide as well the expression for the axial $N \rightarrow \Delta$ 
response, whose relevance for the asymmetry has been already alluded to in the 
Introduction. As an alternative to the approach of \cite{Ama99} we perform 
this task by employing the method of the $\Delta$-hole polarization propagator 
$\Pi^{\mu \nu}$. Specific components of the imaginary part of the latter yield, 
the response function for which we are looking.
The polarization propagator is defined as follows:
\beq
\Pi^{\mu \nu}(q) &=& - i \int \frac{d^4p}{(2 \pi)^4} tr_{spin} tr_{isospin} 
\left[ G(p) \Gamma^{\mu \alpha}_{N\Delta}(q) S_{\alpha \beta}(p+q) 
{\Gamma_{\Delta N}^{\beta \nu}}(q) \right] \nonumber \\
&=& i \int \frac{d^4p}{(2 \pi)^4} \frac{1}{4 E(p) E_\Delta(p+q)} \ 
\frac{1}{p_0 + \omega - E_\Delta(p+q) +i \epsilon} \
\frac{\theta(k_F-|p|)}{p_0-E(p)-i \epsilon} \nonumber \\
&\times& tr\left[ (\rlap/p +m_N) \Gamma^{\mu \alpha}_{N\Delta}(q) 
(\rlap/p + \rlap/q + m_\Delta) \nonumber \right. \\ 
&\times& \left. \left( g_{\alpha \beta}  - \frac{1}{3} \gamma_\alpha 
\gamma_\beta - \frac{2}{3} \frac{(p_\alpha +q_\alpha) (p_\beta+q_\beta)}{\MD^2} - 
\frac{\gamma_\alpha (p_\beta+q_\beta) - 
\gamma_\beta (p_\alpha+q_\alpha)}{3 \MD} \right) 
\gamma^0 {\Gamma^{\beta \nu}}_{N\Delta}^\dagger(q)  \gamma^0\right] {\cal I} \ ,
\label{eq:polariz}
\eeq
where
\beq
\rlap/p &\equiv& \gamma^0 E(p) - \vec{\gamma}\cdot\vec{p}
\eeq
for the nucleon and
\beq
\rlap/p &\equiv& \gamma^0 E_\Delta(p) - \vec{\gamma}\cdot\vec{p} 
\end{eqnarray}
for the $\Delta$, and $\cal I$ is the isospin trace
\begin{eqnarray}
{\cal I} &=& tr_{isospin} T_{N\Delta} T_{N\Delta}^\dagger = \frac{4}{3} \  ,
\end{eqnarray}
$T_{N\Delta}$ being the isospin $N \rightarrow \Delta$ transition operator.
In (\ref{eq:polariz}) the nucleon and the $\Delta$ propagators respectively read
\beq
G(k) &=& \left(\ks + m_N\right) \left[ \frac{1}{k^2-m_N^2+i \epsilon} +
i \frac{\pi}{E(k)} \delta(k_0-E(k)) \theta(k_F-k) \right] \nonumber \\
&=& \frac{1}{2 E(k)} \left\{ (\ks + m_N) \left[ 
\frac{\theta(k-k_F)}{k_0-E(k)+i \epsilon} + 
\frac{\theta(k_F-k)}{k_0-E(k)-i \epsilon} \right] \nonumber \right. \\
&-& \left. \frac{\tilde{\ks}+m_N}{k_0+E(k)- i \epsilon}\right\}
\eeq
and
\beq
S^{\alpha \beta} (k) &=& \meno \frac{\ks + \MD}{k^2-\MD^2+i \epsilon} 
\left( g^{\alpha \beta} - \frac{1}{3} \gamma^\alpha 
\gamma^\beta - \frac{2}{3} \frac{k^\alpha k^\beta}{\MD^2} -
\frac{\gamma^\alpha k^\beta - \gamma^\beta k^\alpha}{3 \MD} \right) 
\nonumber \\
&=&  \meno \frac{1}{2 E_\Delta(k)} \left( \frac{\ks + \MD}{k_0-E_\Delta(k)+
i \epsilon} - 
\frac{\tilde{\ks} + \MD}{k_0+E_\Delta(k)-i \epsilon} \right) \nonumber \\
&\times& \left[ g^{\alpha \beta}  - \frac{1}{3} \gamma^\alpha \gamma^\beta - 
\frac{2}{3} \frac{k^\alpha k^\beta}{\MD^2} -
\frac{\gamma^\alpha k^\beta - \gamma^\beta k^\alpha}{3 \MD} \right] \ ,
\eeq
where $\tilde{\ks} \equiv - \gamma_0 E(k) - \vec{\gamma}\cdot\vec{k}$.

Now the energy integration in (\ref{eq:polariz}) is easily performed, 
yielding
\beq
\Pi^{\mu \nu}(q) &=& - \frac{1}{(2 \pi)^3} \int d^3p 
\frac{m_N^2 f^{\mu \nu}(p,q)}{4 E(p) E_\Delta(p+q)} \ 
\frac{\theta(k_F-|p|)}{\omega+E(p)-E_\Delta(p+q)+i \epsilon}
\label{eq:polariz1}
\eeq
in terms of the dimensionless single-nucleon second-rank $N-\Delta$
tensor
\beq
f^{\mu \nu}(p,q) &\equiv& \frac{1}{m_N^2} tr\left[ (\rlap/p +m_N) 
\Gamma_{N\Delta}^{\mu \alpha}(q) (\rlap/p + \rlap/q +m_\Delta) 
\nonumber \right. \\ 
&& \left. 
 \left( g_{\alpha \beta}  - \frac{1}{3} \gamma_\alpha \gamma_\beta - \frac{2}{3} 
\frac{(p_\alpha +q_\alpha) (p_\beta+q_\beta)}{\MD^2} - \frac{\gamma_\alpha (p_\beta+q_\beta) - 
\gamma_\beta (p_\alpha+q_\alpha)}{3 \MD} \right) \gamma^0 \Gamma^{\nu \beta}_{N\Delta}(q)^\dagger \gamma^0 
\right] \nonumber \ .
\eeq
The latter can be conveniently expressed in terms of the independent tensors
\begin{eqnarray}
\label{eq:xia}
\xi_a^{\mu \nu} &\equiv& g^{\mu \nu} - \frac{q^\mu q^\nu}{q^2} = g^{\mu \nu} + 
\frac{\kappa^\mu \kappa^\nu}{\tau} \\
\label{eq:xib}
\xi_b^{\mu \nu} &\equiv& \frac{1}{m_N^2} \left( p^\mu - \frac{q\cdot p}{q^2} q^\mu \right)
\left( p^\nu - \frac{q\cdot p}{q^2} q^\nu \right) = 
\left( \eta^\mu + \frac{\kappa \cdot \eta}{\tau} \kappa^\mu \right)
\left( \eta^\nu + \frac{\kappa \cdot \eta}{\tau} \kappa^\nu \right) \\ 
\label{eq:xic}
\xi_c^{\mu \nu} &\equiv& i \epsilon^{\mu \nu \alpha \beta}  
\frac{p_\alpha q_\beta}{m_N^2} =  i \epsilon^{\mu \nu \alpha \beta} 
\eta_\alpha \kappa_\beta \ , 
\end{eqnarray}
as follows
\begin{eqnarray}
f^{\mu \nu} =   - w_1 \ \xi_a^{\mu \nu} + w_2 \ \xi_b^{\mu \nu} + 
\tilde{w}_3 \ \xi_c^{\mu \nu} \ ,
\end{eqnarray}
where a tilde has been placed on the term associated with the weak neutral 
current. Note that the $\xi^{\mu \nu}$ are orthogonal to $q^\mu$, i.e. 
$q_{\mu} \ \xi^{\mu \nu}_{a,b,c} = 0$. Also in (\ref{eq:xia}), (\ref{eq:xib}) 
and (\ref{eq:xic}) the two independent four-vectors of our problem, namely 
that of the nucleon inside the FG ($p^\mu$) and the one carried by the 
gauge boson ($q^\mu$), have been expressed in dimensionless forms
according to 
\beq
\eta^\mu &\equiv& ( \epsilon, \vec{\eta}) \equiv \left( \frac{E(p)}{m_N} , \frac{\vec{p}}{m_N} \right)
\eeq
and 
\beq
\kappa^\mu &\equiv& ( \lambda, \vec{\kappa}) \equiv \left( \frac{\omega}{2 m_N} , \frac{\vec{q}}{2 m_N} 
\right) \ ,
\eeq
where furthermore $\tau \equiv \kappa^2 - \lambda^2$.

In order to fix $w_1$, $w_2$ and $\tilde{w}_3$, the vertex functions 
$\Gamma_{N\Delta}^{\mu \alpha}$, whose matrix elements between the 
Rarita-Schwinger spinor $u_\alpha^{(\Delta)}$ and the Dirac nucleon
spinor $u$ yield the $N \rightarrow \Delta$ current, namely
\beq 
\langle \Delta | J^\mu (0) | N \rangle \equiv J^\mu(q) = 
\overline{u}^{(\Delta)}_\alpha(p+q) \Gamma_{\Delta N}^{\alpha \mu} u(p) \ ,
\eeq
are required. These we take from the work of Devenish et al. \cite{Dev76}; they 
have the following expressions
\beq
\label{eq:vert1}
\Gamma_{M(V)}^{\beta \mu} &=& - \frac{3}{2} \frac{\mu + 1}{Q_+} 
\epsilon^{\beta \mu}(p q) \\
\label{eq:vert2}
\Gamma_{E(V)}^{\beta \mu} &=& - \Gamma_{M(V)}^{\beta \mu} -
i \frac{3}{2} \frac{\mu + 1}{Q_+ Q_-} 4 \epsilon^{\beta \sigma}(p q)
\epsilon^\mu_{\sigma}(p q) \ \gamma_5 \\
\label{eq:vert3}
\Gamma_{C(V)}^{\beta \mu} &=& - i \frac{3}{2} \frac{\mu + 1}{Q_+ Q_-}
2 q_\beta \left( q^2 p^\mu - p\cdot q \ q^\mu \right) \ \gamma_5 \\
\label{eq:vert4}
\Gamma_{M(A)}^{\beta \mu} &=& - i \frac{3}{2} \frac{\mu + 1}{Q_-} 
\left[ - 2 i \gamma_5 \epsilon^{\beta \mu}(p q) - \frac{2}{Q_+} 
\epsilon^{\beta \sigma}(p q) \epsilon^{\mu}_\sigma(p q) \right] \\
\label{eq:vert5}
\Gamma_{E(A)}^{\beta \mu} &=& - i \frac{3}{2} \frac{\mu-1}{Q_+ Q_-} 2
\epsilon^{\beta \sigma}(p q) \epsilon^{\mu}_\sigma(p q)  \\
\label{eq:vert6}
\Gamma_{C(A)}^{\beta \mu} &=& i \frac{3}{2} \frac{\mu - 1}{Q_+ Q_-} 2
q^\beta \left( q^2 p^\mu - p\cdot q q^\mu \right) 
\eeq
with $\mu \equiv m_\Delta / m_N$ and 
\beq
Q_\pm &\equiv (m_\Delta \pm m_N)^2 - q_\alpha^2 = m_N^2 \left[ (\mu \pm 1)^2 + 
4 \tau \right] 
\nonumber  \ .
\eeq
In  (\ref{eq:vert1}-\ref{eq:vert6})  the shorthand notation 
$\epsilon^{\beta \sigma}(p q) \equiv \epsilon^{\beta \sigma \mu \nu} p_\mu 
q_\nu$ has been used. 

Then by associating $N \rightarrow \Delta$ form factors, to be denoted 
$G^{(v)}_{E,M,C}(\tau)$ and
$G^{(a)}_{E,M,C}(\tau)$ in the vector and axial sectors respectively,
to each of the vertices  (\ref{eq:vert1}-\ref{eq:vert6}) and by performing the
spin traces, one finally obtains
\beq
w_1(\tau) &=& -\frac{1}{16} \left( 3 G_E^{(v)}(\tau)^2 + G_M^{(v)}(\tau)^2 \right) 
     \left(\mu + 1 \right)^2 \left[ (\mu - 1)^2+ 4 \tau \right] \\
w_2(\tau) &=& \frac{\tau (\mu +1)^2}{(\mu+1)^2+4 \tau} \left[ 
3 G_E^{(v)}(\tau)^2 + G_M^{(v)}(\tau)^2  + \frac{4 \tau}{\mu^2}  
G_C^{(v)}(\tau)^2 \right] \\
\tilde{w}_3(\tau) &=&  \frac{1}{4} (\mu^2-1)  \left( 3 G^{(v)}_{E}(\tau) 
G^{(a)}_{M}(\tau) + G^{(a)}_{E}(\tau) G^{(v)}_{M}(\tau) \right) \ .
\eeq

Worth noticing in the above formula is the disappearance of the axial 
Coulomb multipole which thus does not contribute to the single-nucleon 
$N \rightarrow \Delta$ tensor, as should be the case.

\section{The $N-\Delta$ symmetric RFG responses}

In the previous section we have set up all of the ingredients required to 
compute the $N \rightarrow \Delta$ responses. These obtain, through 
appropriate specifications of the Lorentz indices, according
to the formula
\begin{eqnarray}
{\cal R}^{\mu \nu}(\lambda,\kappa) &=& - \frac{V}{\pi} 
Im \Pi^{\mu \nu}(\lambda,\kappa) = - A  \frac{3 \pi^2}{2 m_N^3 \eta_F^3} \ 
Im\Pi^{\mu \nu}(\lambda,\kappa) \ ,
\label{eq:rmunu}
\end{eqnarray}
where $A$ is the number of nucleons enclosed in a volume $V$ and the 
dimensionless Fermi momentum $\eta_F = k_F/m_N$ has been introduced.

Now if the $\Delta$ is assumed to be a stable particle on its mass-shell 
then the three dimensional integration over the nucleon's momentum in the 
imaginary part of (\ref{eq:polariz1}) can be easily converted
into a  one-dimensional integration over the energy $\epsilon = \sqrt{1+\eta^2}$
by exploiting the energy-conserving delta function.
One thus gets
\begin{eqnarray}
{\cal R}^{\mu \nu}(\lambda,\kappa) &=& - \frac{3}{4} \frac{3 {\cal N}}{4 m_N 
\kappa \eta_F^3} \int^{\epsilon_F}_{\tilde{\gamma}_{-}} d\epsilon 
f^{\mu \nu}(p,q)|_{\theta_0}  \ ,
\label{eq:rmunu1}
\end{eqnarray}
where the angle between $\vec{\eta}$ and $\vec{\kappa}$ in the single-nucleon 
tensor of the $N \rightarrow \Delta$ sector is fixed by the energy conservation 
to be
\beq
\cos\theta_0 &=& \frac{ \lambda \epsilon - \tau \rho}{\kappa \eta} \ ,
\eeq
with
\beq
\rho = 1 + \frac{\mu^2-1}{4 \tau}  \ . 
\eeq
Moreover the upper limit of integration in (\ref{eq:rmunu})  is set by the 
dimensionless Fermi energy $\epsilon_F = \sqrt{1+\eta_F^2}$, whereas the 
lower limit 
\beq 
\tilde{\gamma}_{-} \equiv \kappa \sqrt{\frac{1}{\tau}+\rho^2}-\lambda \rho 
\label{eq:gammam}
\eeq
extends to the $\Delta$ domain the $\gamma_-$ of the QEP sector \cite{Alb95} 
to which indeed it reduces when $\rho \rightarrow 1$ ($\mu \rightarrow 1$).

The energy integration in  (\ref{eq:rmunu1}) can be carried out without difficulty
and one gets for the longitudinal, transverse and the axial channels the expressions
\beq
\label{eq:RL}
R^L(\kappa, \lambda) &=& C \int_{\tilde{\gamma_-}}^{\epsilon_F} \ d\epsilon \ 
f^{00}(\epsilon, \theta_0) = C \  \xi_F \ (1-\psi^2) \ U_L(\kappa,\lambda) \\
\label{eq:RT}
R^T(\kappa, \lambda) &=& C \int_{\tilde{\gamma_-}}^{\epsilon_F} \ d\epsilon \ 
\left( f^{11}(\epsilon, \theta_0) + f^{22}(\epsilon, \theta_0) \right) = 
C \  \xi_F \ (1-\psi^2) \ U_T(\kappa,\lambda) \\
\label{eq:RT1}
\tilde{R}^{T'}_{VA}(\kappa, \lambda) &=& - i C 
\int_{\tilde{\gamma_-}}^{\epsilon_F} \ d\epsilon \ 
f^{12}(\epsilon, \theta_0) = C \  \xi_F \ (1-\psi^2) \ U_{T'}(\kappa,\lambda) \ ,
\eeq
where 
\beq
C = \frac{3 \ A}{4 m_N \kappa \eta_F^3} 
\eeq
and the indices on the left hand side of  (\ref{eq:RT1}) refer to the 
coupling of the vector current of the lepton with the axial current of the hadron.

The above responses, just as happens in the QEP domain \cite{Don92}, display 
a common factor $C \xi_F (1-\psi^2)$ and therefore scale, i.e., 
depend only  upon a single variable 
\beq
\psi = \sqrt{\frac{1}{\xi_F} \left( \kappa \sqrt{\frac{1}{\tau}+\rho^2}-
\lambda \rho -1 \right)} \ \ \  \left\{ 
\begin{array}{c} 
+ 1 \ \ , \ \ \ \lambda > \lambda_0 \\
- 1 \ \ , \ \ \ \lambda < \lambda_0 
\end{array} \right.
\label{eq:psi}
\eeq
where 
\beq
\lambda_0 = \frac{1}{2} \left( \sqrt{1 + 4 \kappa^2 \rho} - 1 \right) \ . 
\eeq
Again (\ref{eq:psi}) reduces to the scaling variable of the QEP domain when
$\rho \rightarrow 1$ ($\mu \rightarrow 1$) and carries the physical 
significance of the minimum energy (in units of the Fermi kinetic energy $\xi_F =
\epsilon_F -1$) a nucleon should have to respond to the external field. 

The common factor discussed above reflects the many-body physics of the RFG. 
The $U$ factors in (\ref{eq:RL}), (\ref{eq:RT}) and (\ref{eq:RT1}) relate 
instead mostly (but not only)  to the single-nucleon physics. They read
\beq
U_L(\kappa,\lambda) &=& \frac{\kappa^2}{\tau} \left[ (1+\tau \rho^2) \ w_2(\tau) - 
w_1(\tau) + w_2(\tau) \ {\cal D}^{L}(\kappa,\lambda) \right] \\
U_T(\kappa,\lambda) &=& 2  \ w_1(\tau) + w_2(\tau) \ 
{\cal D}^{T}(\kappa,\lambda)  \\
\tilde{U}_{T'}(\kappa,\lambda) &=& 2  \sqrt{\tau (1+\tau \rho^2)} \ 
\tilde{w}_3(\tau) \left[ 1 +  {\cal D}^{T'}(\kappa,\lambda) \right]  
\eeq
and feel the impact of the medium on the single-nucleon physics through the 
quantities ${\cal D}^{L}$, ${\cal D}^{T}$ and ${\cal D}^{T'}$. Indeed the 
longitudinal ${\cal D}^L$ and the transverse ${\cal D}^T$ simply express the 
mean square value of the nucleon transverse momentum in the medium, i.e.
\beq
{\cal D}^L(\kappa,\lambda) &=& {\cal D}^T(\kappa,\lambda) = 
\frac{1}{\epsilon_F -\tilde{\gamma}_-} \int_{ \tilde{\gamma}_-}^{\epsilon_F} \ 
d\epsilon \ \eta_\perp^2 \nonumber \\
&=& \frac{\tau}{\kappa^2} \left[ (\lambda \rho +1)^2 + (\lambda \rho +1) \ 
(1+\psi^2) \ \xi_F + \frac{1}{3} (1 + \psi^2 + \psi^4) \ \xi_F^2\right] - 
(1 + \tau \rho^2) \ , 
\eeq
whereas the axial ${\cal D}^{T'}$ turns out to be closely related to the mean 
transverse kinetic energy of the nucleon inside the RFG according to
\beq
{\cal D}^{T'} &=&  \frac{1}{\epsilon_F-\tilde{\gamma}_-} 
\int_{\tilde{\gamma}_-}^{\epsilon_F} \ d\epsilon \ \left( \sqrt{1+ 
\frac{\eta_\perp^2}{1+\tau \rho^2}}-1 \right) \nonumber \\
&=& \frac{1}{\kappa} \sqrt{\frac{\tau}{1+\tau \rho^2}} \left[ 1 + \xi_F 
(1+\psi^2) + \lambda \rho \right] - 1 \ .
\eeq

We refer the reader to \cite{Alv00} for a more thorough discussion of the 
${\cal D}$'s : here we simply recall that they vanish, as should be the case, 
when $\eta_F \rightarrow 0$  and that for $\rho \rightarrow  1$  they evolve 
into the corresponding quantities for the QEP \cite{Amo97}.

From the above discussion it thus emerges that the form factors substitutions 
\beq
3 G_E^{(v)}(\tau)^2 + G_M^{(v)}(\tau)^2  \rightarrow G_M(\tau)^2
\label{eq:sub1}
\eeq
and 
\beq
(2 \tau G_C^{(v)}(\tau))^2 \rightarrow G_E(\tau)^2 \ , 
\label{eq:sub2}
\eeq
with $G_{E,M}(\tau)$ being the nucleon's electric and magnetic form factors, 
allow one naturally to recover the RFG responses in the QEP  from those in the
$N \rightarrow \Delta$ domain by letting $\rho \rightarrow 1$. 
Of course in carrying out this procedure the factor $(\mu^2-1)$ in 
$\tilde{w}_3(\tau)$ should be disregarded.

This supports the feasibility of devising a universal dividing factor 
as was done in the QEP to {\sl reduce} the nuclear responses over an energy 
range encompassing an extended region of the spectrum of nucleon 
excitation by exploiting 
form factors substitutions of the type (\ref{eq:sub1}) and (\ref{eq:sub2}). 
This might help when attempting to analyze nuclear $y$-scaling at
higher energies.

As mentioned in the Introduction
in order to compute the RFG $N \rightarrow \Delta$ responses we still have to 
face the issue of the  $N \rightarrow \Delta$ form factors. Here we 
simply provide the expressions we have employed 
in our calculation. They are
\begin{eqnarray}
\label{eq:GMV}
G_{M}^{(v)}(\tau) &=& \frac{G_M^{(v)}(0)}{(1+\lambda_M^V \tau)^2} 
\frac{1}{\sqrt{1+ \tau}} \\
\label{eq:GEV}
G_{E}^{(v)}(\tau) &=& \frac{G_E^{(v)}(0)}{(1+\lambda_E^V \tau)^2} 
\frac{1}{\sqrt{1+ \tau}} \\
\label{eq:GCV}
G_{C}^{(v)}(\tau) &=& \frac{G_C^{(v)}(0)}{(1+\lambda_C^V \tau)^2} 
\frac{1}{\sqrt{1+ \tau}} 
\eeq
in the vector sector and
\beq
G_{M}^{(a)}(\tau) &=& \frac{G_M^{(a)}(0)}{(1+\lambda_M^A \tau)^2}  \\
G_{E}^{(a)}(\tau) &=& \frac{G_E^{(a)}(0)}{(1+\lambda_E^A \tau)^2}  \\
G_{C}^{(a)}(\tau) &=& \frac{G_C^{(a)}(0)}{(1+\lambda_C^A \tau)^2}  
\eeq
in the axial one. Below we quote the values of the parameters entering 
into the above formulas used here:
\beq
G_M^{(v)}(0) &=& 2.97 \ \ \ , \ \ \ G_E^{(v)}(0) = -0.03 \ \ \ , \ \ \
G_C^{(v)}(0) = -0.44 \\
G_M^{(a)}(0) &=& 0 \ \ \ \ \ \ , \ \ \ G_E^{(a)}(0) = 2.22 \ \ \ \ \ \ , \ \ \
G_C^{(a)}(0) = 0 \\
\lambda_M^V &=& 4.97 \ \ \ , \ \ \ \lambda_E^V = 4.97 \ \ \ , \ \ \
\lambda_C^V = 4.97  \\
\lambda_M^A &=& 3.53 \ \ \ , \ \ \ \lambda_E^A = 3.53 \ \ \ , \ \ \
\lambda_C^A = 3.53   
\end{eqnarray}

\begin{figure}
\begin{center}
\mbox{\epsfig{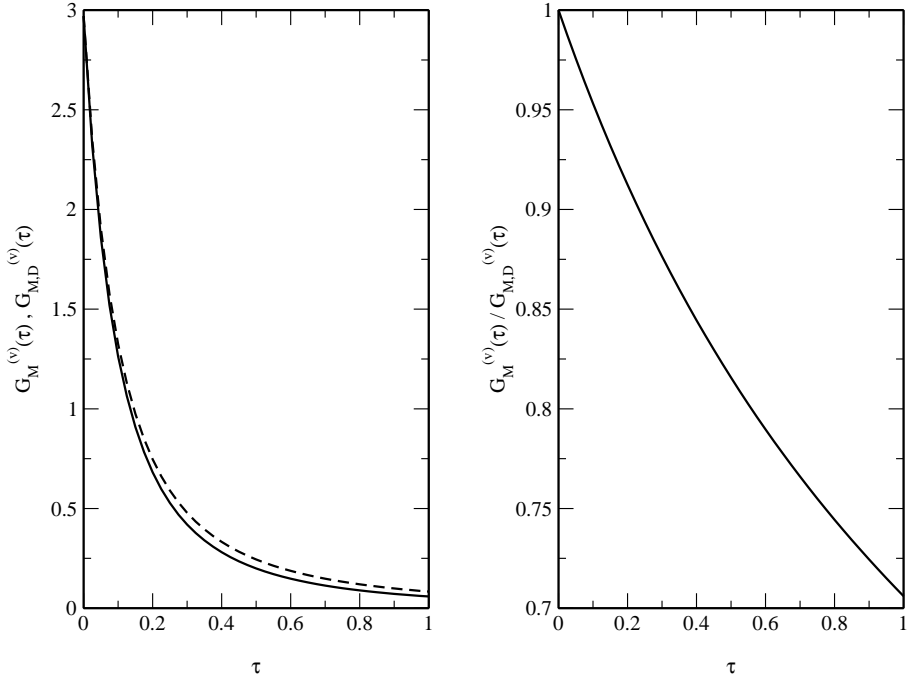}}
\end{center}
\caption{Comparison of the $G_M^{(v)}(\tau)$ with the pure dipole form 
(dashed curve).  Their ratio is displayed in the right panel.}
\label{fig:1}
\end{figure}

A few comments are worth making at this junction:
\begin{enumerate}
\item in the axial sector we only retain the electric $N \rightarrow \Delta$ 
form factor, the largest one according to the constituent quark model 
\cite{Hem95};

\item its value at the origin has been fixed in accord with a scaling law
\cite{Adl68};

\item the square root in the denominator of (\ref{eq:GMV}), (\ref{eq:GEV}) and 
(\ref{eq:GCV}) yields a decrease of about $30 \%$ at  $\tau = 1$  of 
the vector form factors with respect to pure dipole behavior 
(see Fig. \ref{fig:1}).

\end{enumerate}

We are now ready to display our results for both the PC and PV RFG responses
in the nucleon and $\Delta$ sectors. We shall consider three 
different kinematical domains. Indeed it is well-known that 
the quasi-elastic and the $\Delta$ responses 
occur in the $(\lambda, \kappa)$ plane in the domains defined by the curves
\beq
\lambda_{1,2}^{(N)} = \frac{1}{2} \left\{ \sqrt{(2 \kappa \pm \eta_F)^2+1} - \sqrt{1+\eta_F^2} \right\}
\label{eq:regionN}
\eeq
and
\beq
\lambda_{1,2}^{(\Delta)} = \frac{1}{2} \left\{ \sqrt{(2 \kappa \pm \eta_F)^2+\mu^2} - 
\sqrt{1+\eta_F^2} \right\} \ .
\label{eq:regionD}
\eeq

\begin{figure}
\begin{center}
\mbox{\epsfig{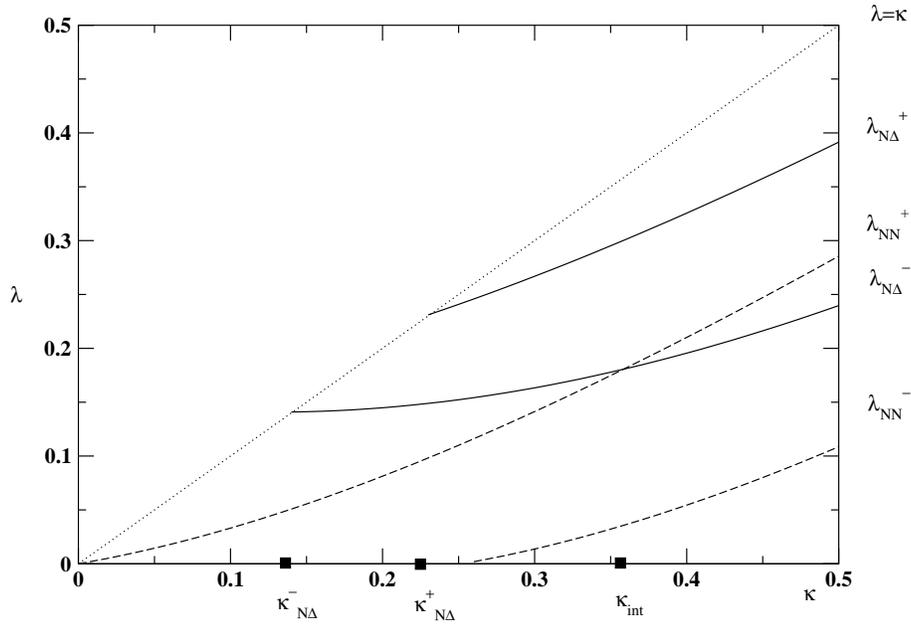}}
\end{center}
\caption{The $N \rightarrow \Delta$ and $N \rightarrow N$ response regions in the  
$(\lambda,\kappa)$ plane for $\eta_F = 0.28$. For $\kappa \leq \kappa_{N\Delta}^{(-)}$ the 
$\Delta$ cannot be excited by space-like photons (the response region lies entirely 
in the time-like domain); for $\kappa_{N\Delta}^{(-)} \leq \kappa \leq \kappa_{N\Delta}^{(+)}$
only part of the response region is accessible (the light-cone lies inside
the response region); for $\kappa_{N\Delta}^{(+)} \leq \kappa \leq 
\kappa_{int}$ the response regions of the $\Delta$ and nucleon 
are still separated; for $\kappa \geq \kappa_{int}$ the response regions of 
the $\Delta$ and nucleon overlap.}
\label{fig:2}
\end{figure}

These are displayed in Fig. \ref{fig:2} together with the light-front.
From the figure the three domains referred to above are apparent, namely:
\begin{enumerate}
\item the region where the $N$ and $\Delta$ response region do not overlap 
and the $\Delta$ region is cut by the light-front, occurring  for 
$\kappa_{N\Delta}^{-} \leq \kappa \leq \kappa_{N\Delta}^{+}$, where 
\beq
\kappa_{N\Delta}^{\pm} = \frac{\mu^2-1}{4} (\epsilon_F \pm \eta_F) \ ;
\eeq

\item the region where the two domains do not overlap  and the
$\Delta$ region is not cut by the light-front, occurring for 
$\kappa_{N\Delta}^{+} \leq \kappa \leq \kappa_{int}$, where 
\beq
\kappa_{int} = \frac{\mu^2-1}{8 \eta_F} \ ;
\eeq

\item the region where the two domains overlap, with both lying below the 
light-front, occurring for $\kappa \geq \kappa_{int}$.
\end{enumerate}

\begin{figure}
\begin{center}
\mbox{\epsfig{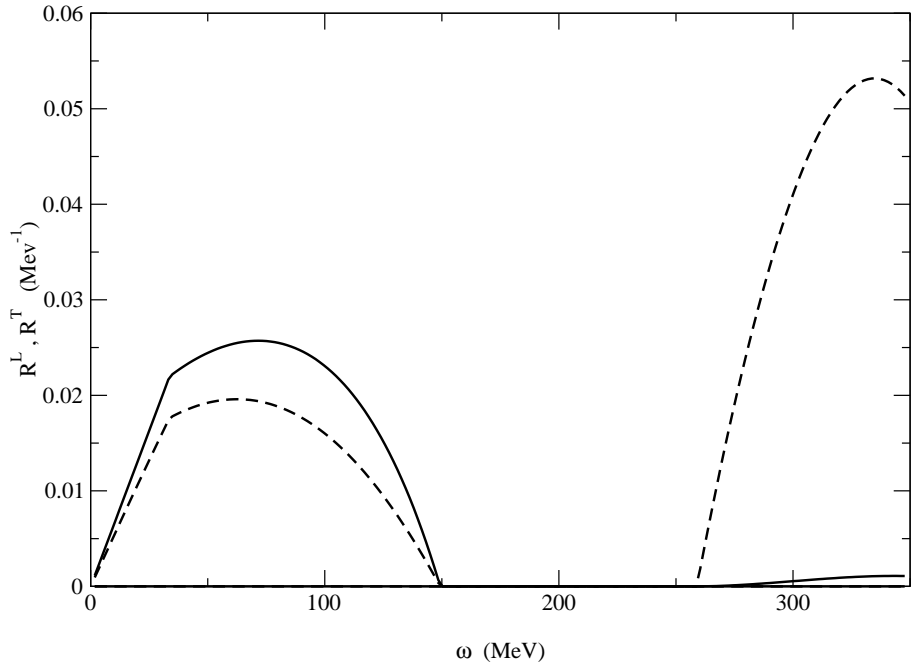}}
\end{center}
\caption{Longitudinal (solid) and transverse (dashed) RFG 
responses at $q=350 \ {\rm MeV/c}$. The Fermi momentum is $k_F = 220 \ {\rm MeV/c}$, here  
and in all the following figures.}
\label{fig:3}
\end{figure}

\begin{figure}
\begin{center}
\mbox{\epsfig{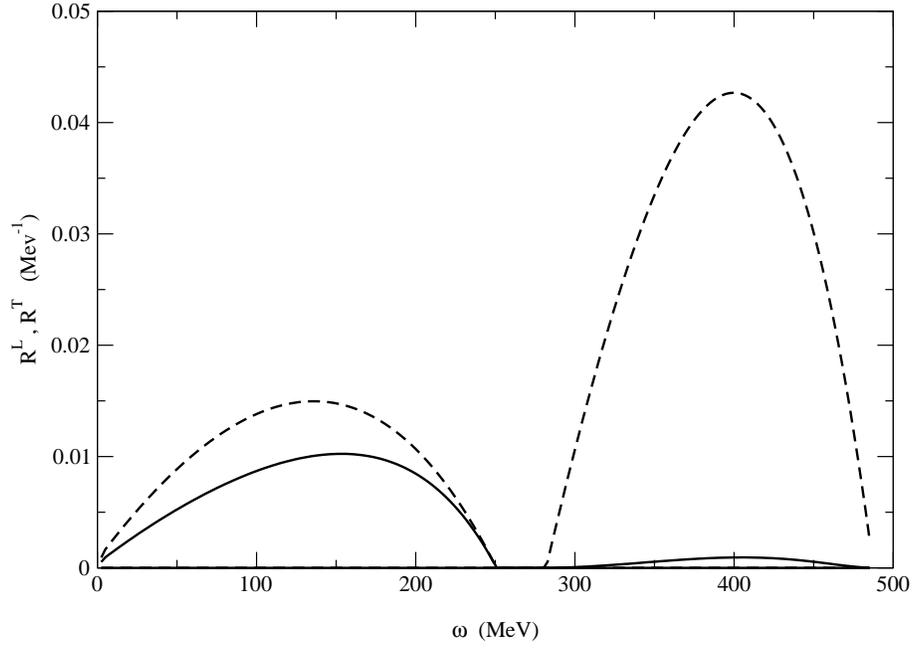}}
\end{center}
\caption{Longitudinal (solid) and transverse (dashed) RFG responses at $q=520 \ {\rm MeV/c}$.}
\label{fig:4}
\end{figure}

\begin{figure}
\begin{center}
\mbox{\epsfig{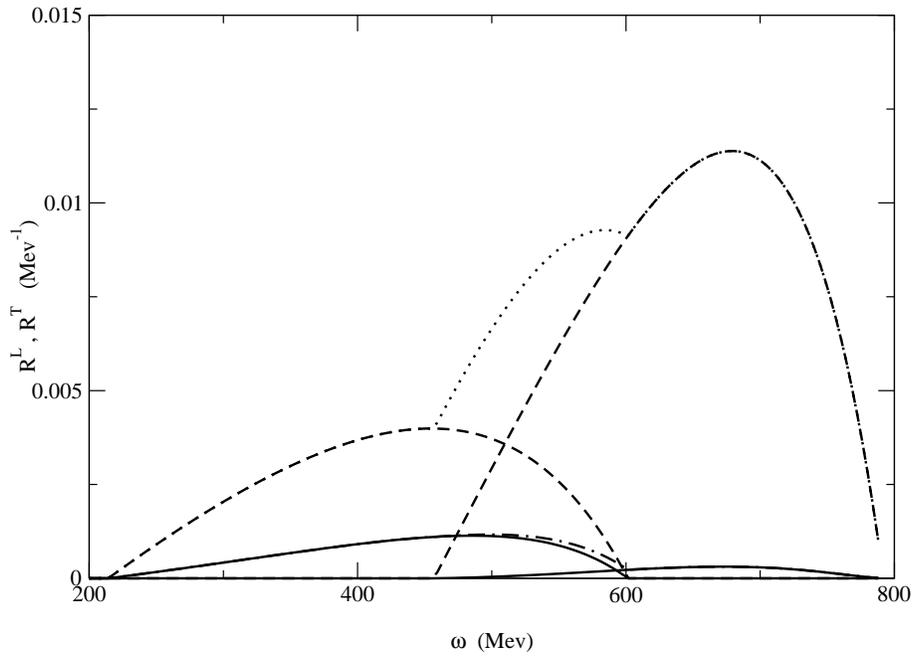}}
\end{center}
\caption{Longitudinal (solid) and transverse (dashed) RFG responses at $q=1000 \ {\rm MeV/c}$.
The dotted lines represent the total of the $\Delta$ and nucleon contributions.}
\label{fig:5}
\end{figure}

Our results for the PC responses are displayed in Figs. \ref{fig:3}, 
\ref{fig:4} and \ref{fig:5} for $k_F = 220 \ {\rm MeV/c}$, roughly 
corresponding to the density of $^{12}$C.
Henceforth we use $q = 350, \ 520$ and $ 1000 \ {\rm MeV/c}$, which are
representative of the three regions referred to above. 
One sees that the transverse contribution in the $\Delta$ region is dominant in
all three cases considered. 
In the QEP the transverse contribution gradually takes over the 
longitudinal one as $q$ increases. Finally at $q  = 350 \ {\rm MeV/c}$ 
the linear behavior of the QEP response functions for small $\omega$ 
reflects the Pauli blocking.

\begin{figure}
\begin{center}
\mbox{\epsfig{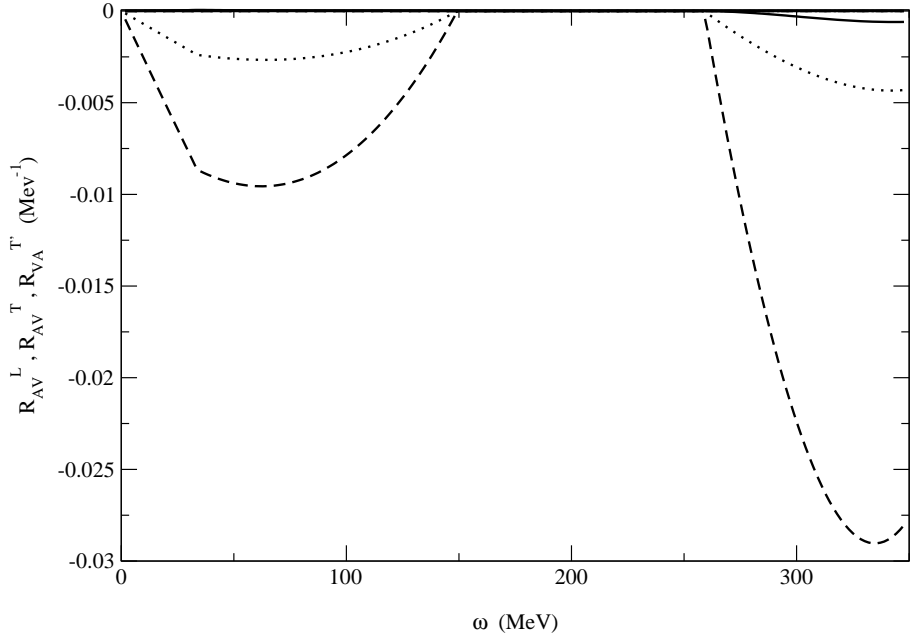}}
\end{center}
\caption{The PV $\tilde{R}^L_{AV}$ (solid), $\tilde{R}^T_{AV}$ (dashed) and $\tilde{R}^{T'}_{VA}$ 
(dotted) at $q=350 \ {\rm MeV/c}$.}
\label{fig:6}
\end{figure}

\begin{figure}
\begin{center}
\mbox{\epsfig{file=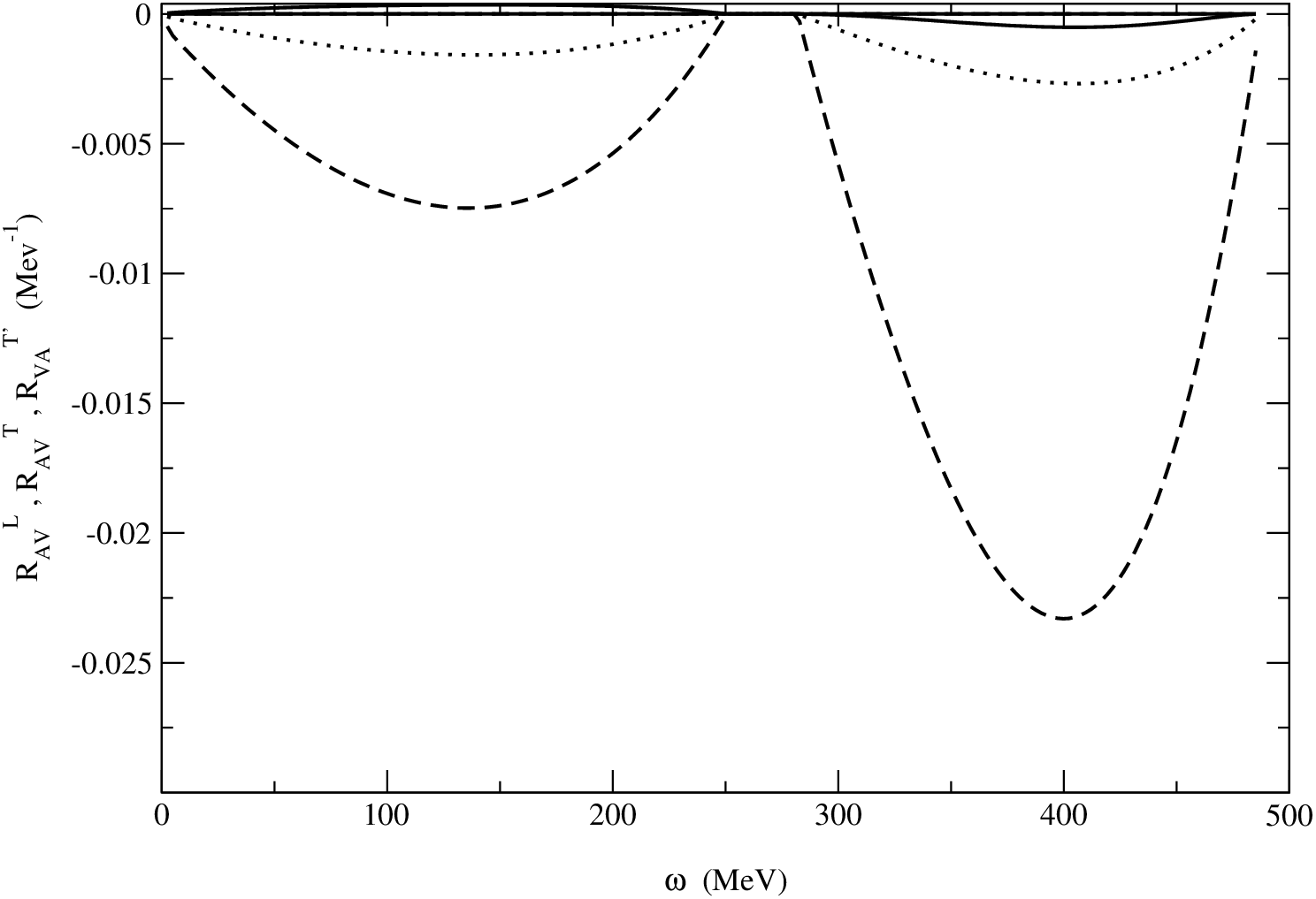, width=12cm}}
\end{center}
\caption{The PV $\tilde{R}^L_{AV}$ (solid), $\tilde{R}^T_{AV}$ (dashed) and $\tilde{R}^{T'}_{VA}$ 
(dotted) at $q=520 \ {\rm MeV/c}$.}
\label{fig:7}
\end{figure}

\begin{figure}
\begin{center}
\mbox{\epsfig{file=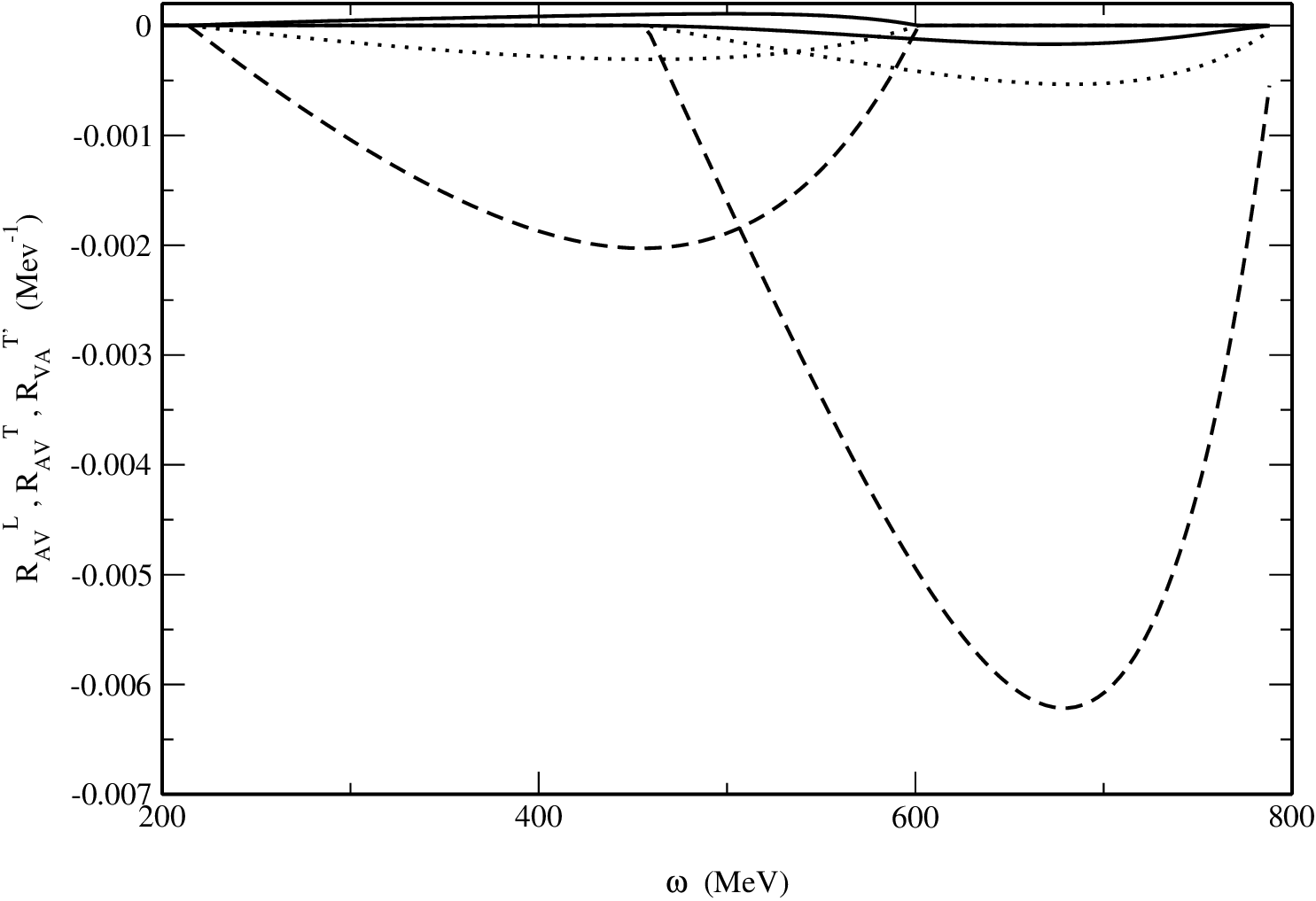, width=12cm}}
\end{center}
\caption{The PV $\tilde{R}^L_{AV}$ (solid), $\tilde{R}^T_{AV}$ (dashed) and $\tilde{R}^{T'}_{VA}$ 
(dotted) at $q=1000 \ {\rm MeV/c}$.}
\label{fig:8}
\end{figure}

We now turn to a consideration of the PV responses. 
The longitudinal and transverse ones are obtained  according to
\beq
\tilde{R}_{AV}^{L,T} &=& \beta^{I=0} \ R_{L,T}^{N}(I=0) + 
\beta^{I=1} \ R_{L,T}^{N}(I=1) \ ,
\label{eq:rot}
\eeq
where 
\beq
\beta^{I=0} = - 2 \sin^2\theta_w \ \ \  , \ \ \ \beta^{I=1} = 
( 1 - 2 \sin^2\theta_w )  \ , 
\label{eq:beta}
\eeq 
$I$ is the isospin quantum number and, as in (\ref{eq:RT1}), the indices refer to
the coupling of the axial current of the lepton with the vector current of the 
hadron. These are displayed in Fig. \ref{fig:6}, \ref{fig:7} and 
\ref{fig:8}. From the figures the dominance of the transverse contribution 
in the $\Delta$ sector is again apparent, just as in the PC case. 
Now, however, the axial contribution is appreciable, the more so the smaller is 
$q$. The same happens in the QEP.
The longitudinal channel turns out to be small in both sectors, but for
different reasons. In the $\Delta$ sector $\tilde{R}_{AV}^L$ is small because so
is $G_C^{(v)}(\tau)$ at moderate values of $\tau$, which likewise also 
implies that the magnetic contribution to the longitudinal channel is small. 
In the QEP sector instead the smallness of $R^L$ stems from the competition 
between its isoscalar and isovector components implied by (\ref{eq:rot}) 
and (\ref{eq:beta}). The same argument clearly does not apply to $R^T$ because 
in this case the isoscalar contribution is quenched by the smallness of the 
isoscalar magnetic moment of the nucleon.

\begin{figure}
\begin{center}
\mbox{\epsfig{file=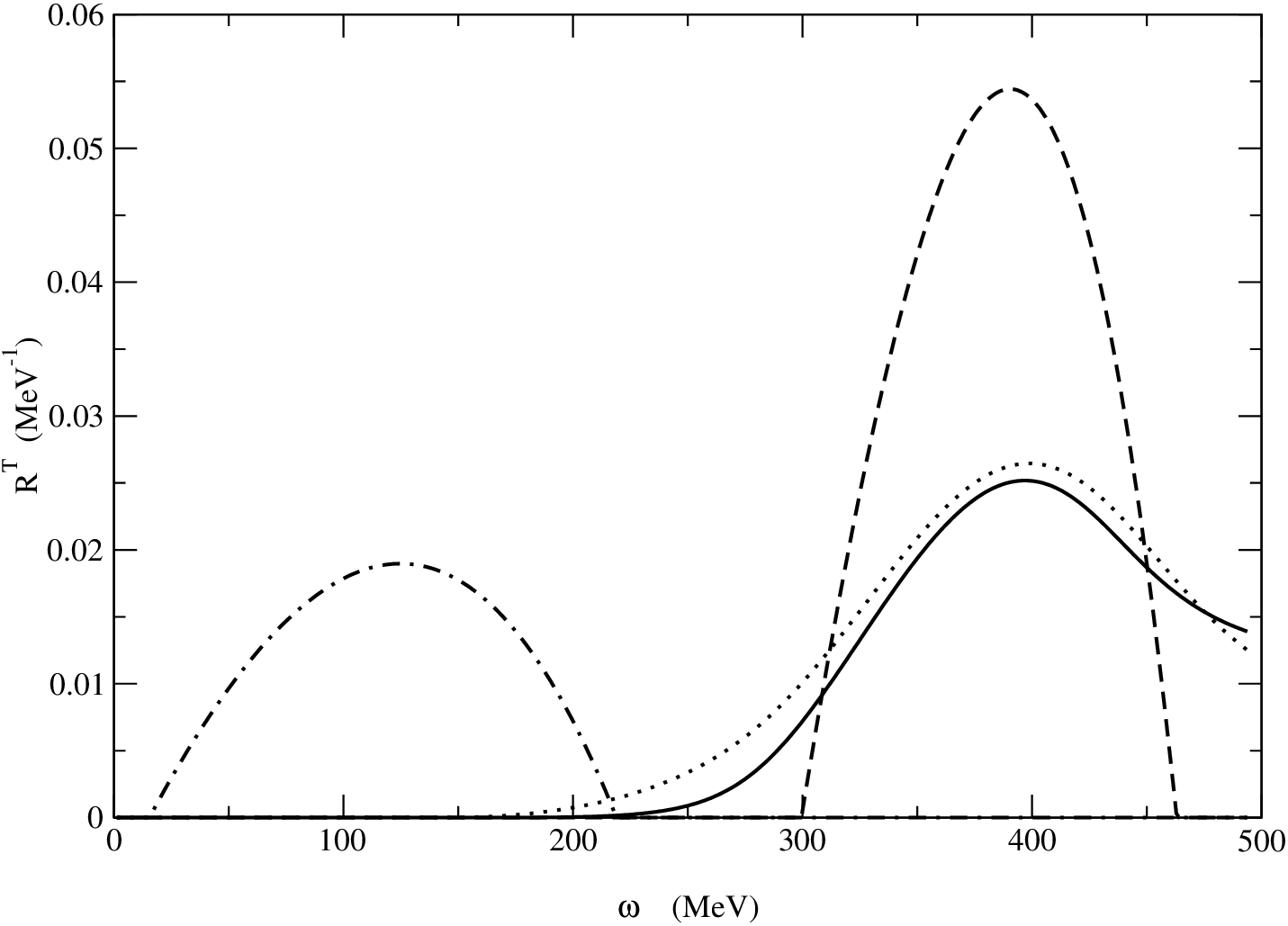, width=12cm}}
\end{center}
\caption{The transverse response is displayed at $q = 500 \ {\rm MeV/c}$ in the nucleonic 
(dot-dashed line) and $\Delta$ sector. In the latter the dashed curve corresponds
to zero width for the $\Delta$, the dotted line to a constant width of 
$\Gamma = 120 \ {\rm MeV}$  and the solid line to an energy-dependent width $\Gamma(s)$.
}
\label{fig:9}
\end{figure}

\begin{figure}
\begin{center}
\mbox{\epsfig{file=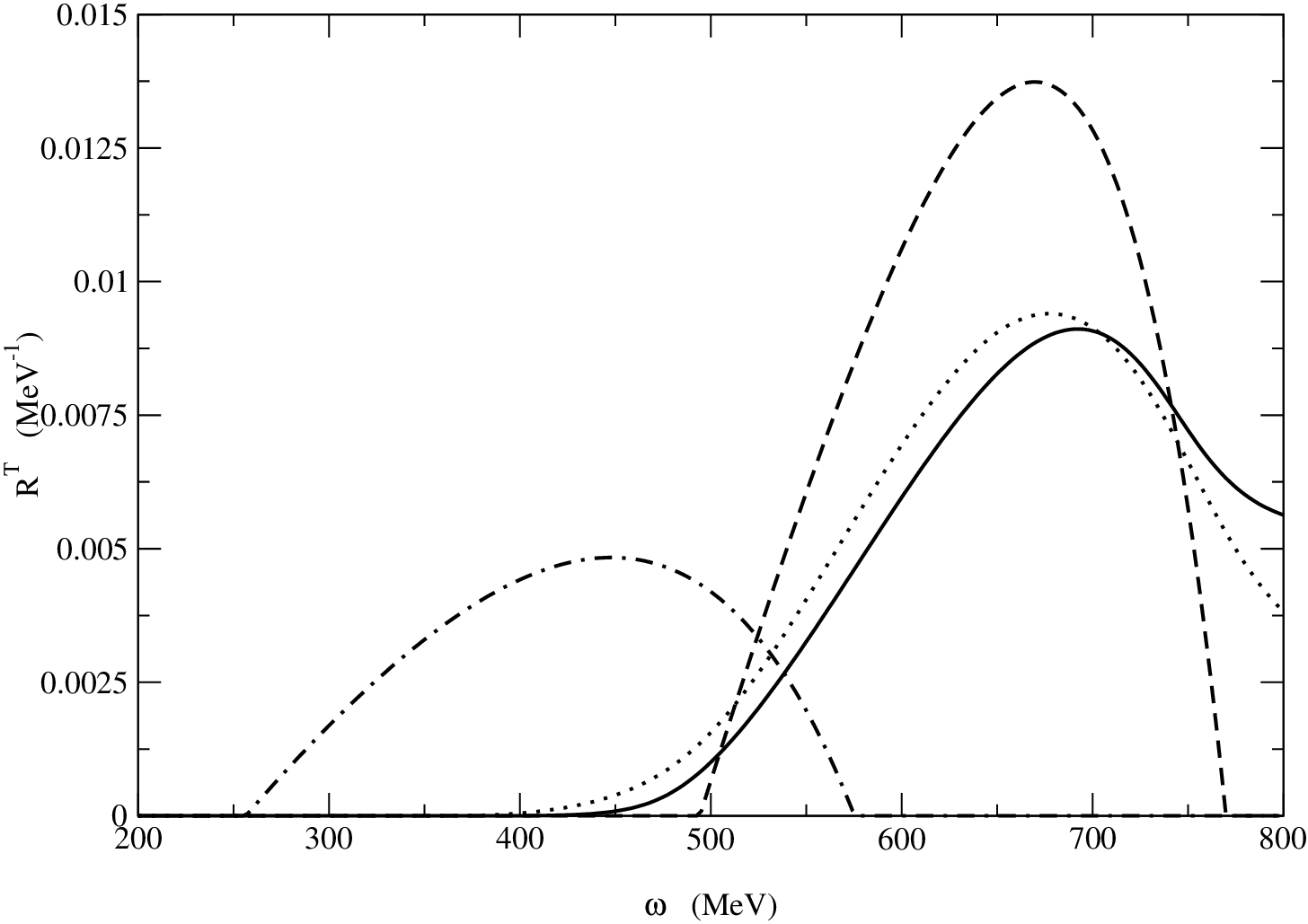, width=12cm}}
\end{center}
\caption{The transverse response is displayed at $q = 1000 \ {\rm MeV/c}$ in the nucleonic 
(dot-dashed line) and $\Delta$ sector. In the latter the dashed curve corresponds
to zero width for the $\Delta$, the dotted line to a constant width of 
$\Gamma = 120 \ {\rm MeV}$ and the solid line  to an energy-dependent width $\Gamma (s)$.}
\label{fig:10}
\end{figure}

The responses computed in this section have so far been obtained viewing the 
$\Delta$ as an elementary particle. We evaluate the impact of the  width of 
the $\Delta$ by following \cite{Ama99}: we write
\begin{equation}
R_\Gamma(q,\omega)= \int_{m_N+m_\pi}^{W_{max}} 
\frac{1}{\pi}\frac{\Gamma(W)/2}{(W-m_\Delta)^2+\Gamma(W)^2/4}
R(q,\omega,W)dW \ ,
\end{equation}
where the integration interval goes from threshold to the maximum value 
allowed in the Fermi gas model, i.e. $W_{max}^2 = (E_F+\omega)^2-(q-k_F)^2$. 
Our results are displayed in Figs. \ref{fig:9} and \ref{fig:10} 
for $q = 0.5 \ {\rm GeV/c}$  and $q = 1 \ {\rm GeV/c}$
and are obtained  both with a constant width $\Gamma = 120 \ {\rm MeV}$ and 
with an energy-dependent width, taken from \cite{Ama99}. As expected the 
inclusion of the $\Delta$ width produces a broadening and, correspondingly, 
a decrease of the strength of the $\Delta$ peak, the more so the smaller is
the momentum transfer. 
The energy dependence of $\Gamma$ is seen to have a modest impact 
and clearly it yields results quite close to those obtained with a constant $\Gamma$.

\section{Asymmetry}

In this section we compute the asymmetry of the symmetric RFG on the basis of the response functions
obtained in the previous sections. In the one gauge boson ($\gamma$ or $Z_0$) exchange approximation
the asymmetry, namely the ratio between the inclusive, inelastic PV and PC cross sections,
reads 
\beq
{\cal A} &=& \frac{\left( \frac{d^2\sigma}{d\Omega \ d\epsilon'}\right)^{(PV)}}{\left( \frac{d^2\sigma}{d\Omega \ 
d\epsilon'}\right)^{(PC)}} =  {\cal A}_0 \ \frac{v_L \tilde{R}_{AV}^L(q,\omega) + v_T \tilde{R}_{AV}^T(q,\omega) 
+v_{T'} \tilde{R}_{VA}^{T'}(q,\omega)}{v_L R^L(q,\omega) + v_T R^T(q,\omega)} \ ,
\label{eq:asym}
\eeq
where
\beq
{\cal A}_0 &=& \frac{\sqrt{2}}{\pi \alpha} \ \left( G m_N^2 \right) \tau \approx 6.5 \ 10^{-4} \ ,
\eeq
$\alpha$ and $G$ are the fine structure and Fermi coupling constants and 
\beq
v_L = \left(\frac{\tau}{\kappa^2}\right)^2 \ \ \ ,  \ \ \ v_T =  \frac{1}{2} \frac{\tau}{\kappa^2} +
\tan^2 \frac{\theta}{2} \ \ \  , \ \ \ v_{T'} = \tan \frac{\theta}{2} \ \sqrt{ \frac{\tau}{\kappa^2}
+  \tan^2 \frac{\theta}{2}} \ .
\label{eq:vl}
\eeq
The responses appearing in (\ref{eq:asym}) are generic. If, however, we restrict 
(\ref{eq:asym}) to the domain of the $\Delta$, which is a {\sl pure isovector} 
excitation of the nucleon, then the asymmetry becomes
\beq
{\cal A}_{N-\Delta} &=& {\cal A}_0 \left\{ - \left(1-2 \sin^2\theta_w\right) + 
v_{T'} \ \frac{\tilde{R}_{VA}^{T'}(q,\omega)}{v_L R^L(q,\omega) + 
v_T R^T(q,\omega)} \right\} \ . 
\label{eq:asymnd}
\eeq

The above formula clearly shows that if the axial $N \rightarrow \Delta$ 
response can be neglected then the 
inelastic asymmetry, normalized to ${\cal A}_0$ and displayed versus $\lambda$ for fixed $\kappa$, would be
flat in the $\Delta$ domain. Hence a departure from flatness would signal the presence of the
axial response. The contribution of the latter is however suppressed with respect to the first 
term on the right hand side of (\ref{eq:asymnd}) by the smallness of 
the vector coupling of the lepton to
the axial current of the hadron, whose value is  $(-1 + 4 \sin^2\theta_w) \approx - 0.092$.

It should, however, be observed that, unlike the case of parity-violating elastic electron
scattering, here the axial contribution to the asymmetry does not vanish at forward electron scattering
angles. Indeed (see \cite{Muk98} for details) from (\ref{eq:vl})  it follows that
when $\theta \rightarrow 0$ and $\tau \rightarrow 0$ one has
\beq
\frac{v_L}{v_T} \rightarrow 0 \ ,
\eeq 
but 
\beq
\frac{v_{T'}}{v_T} \rightarrow \frac{\epsilon^2-{\epsilon'}^2}{\epsilon^2+{\epsilon'}^2} \neq 0 \ ,
\eeq 
where $\epsilon$ and $\epsilon'$ are the initial and final electron energies, 
respectively.

We thus {\sl a priori} expect that, small as it might be,  the best possibility of detecting the
axial response in the $\Delta$ region should be found for not too large momentum transfers
and as near as possible to the light cone. 

\begin{figure}
\begin{center}
\mbox{\epsfig{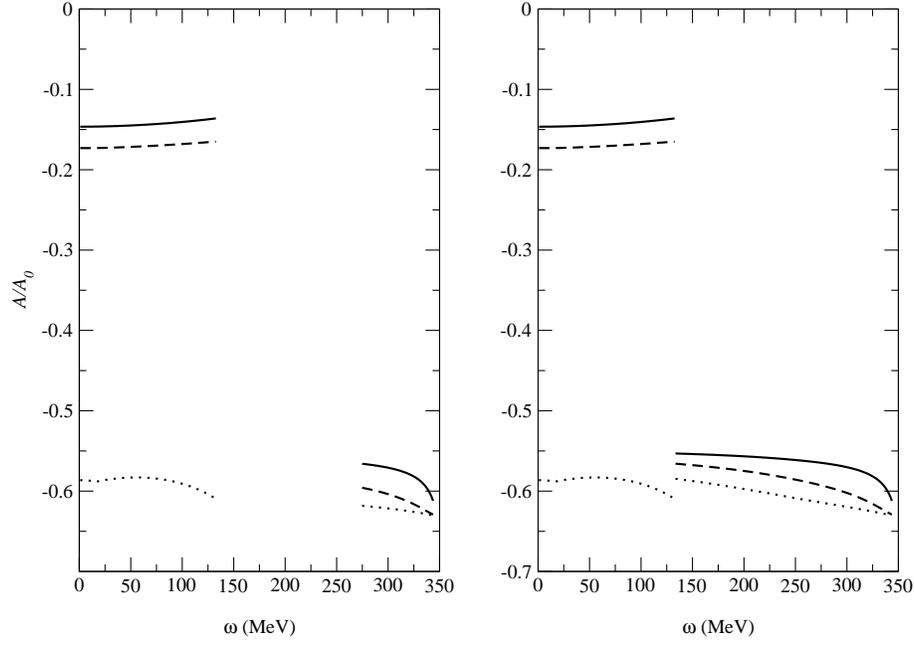}}
\end{center}
\caption{Asymmetry at $\theta=10^0$ (solid), $\theta=30^0$ (dashed) and
$\theta=150^0$ (dotted) for $q=350 \ {\rm MeV/c}$. The $\omega$ range encompasses both
the QEP and the $\Delta$ domain. The left and right panels respectively refer 
to a vanishing width and to  a finite decay width $\Gamma(s)$.}
\label{fig:asym1new}
\end{figure}

\begin{figure}
\begin{center}
\mbox{\epsfig{file=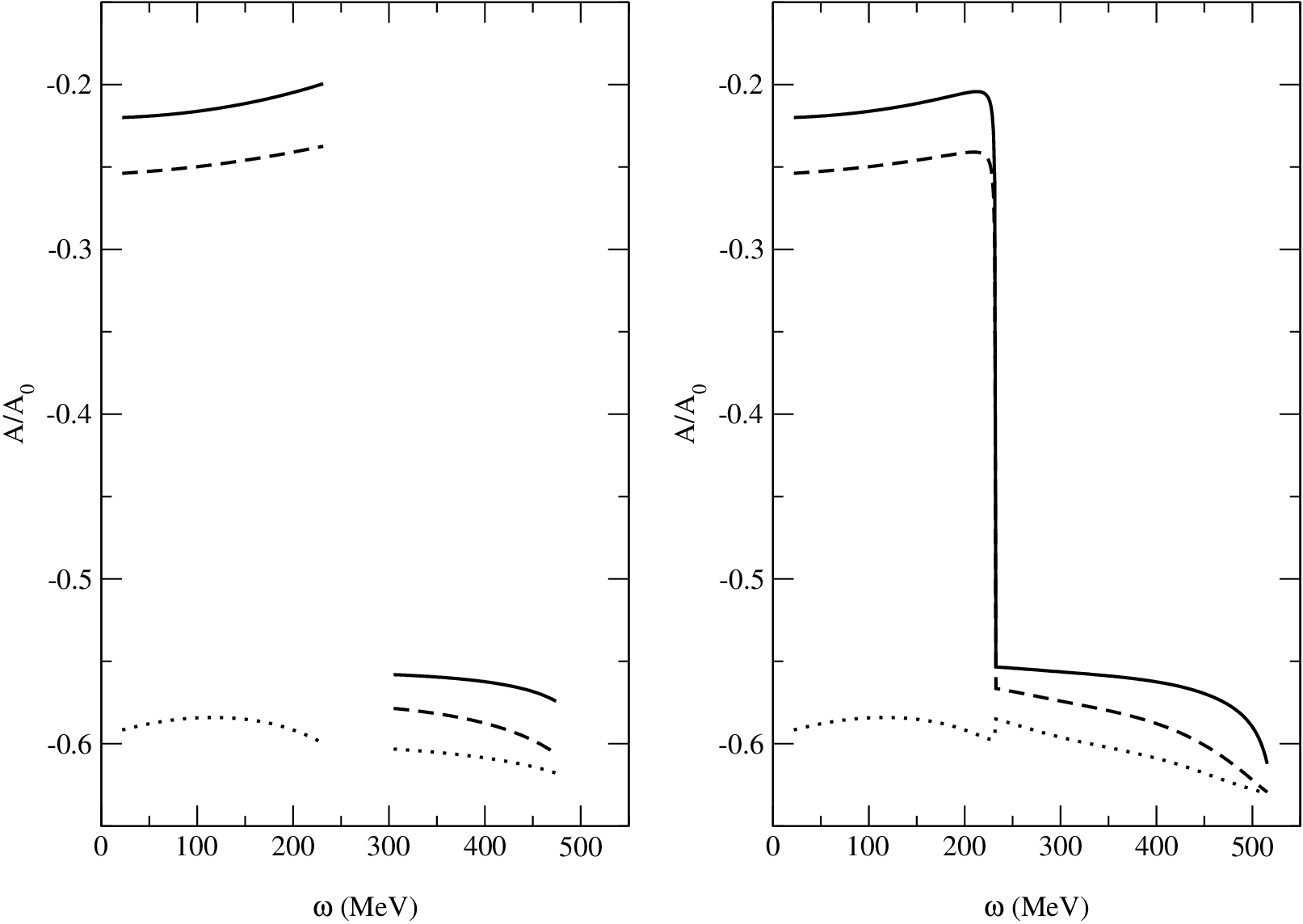, width=12cm}}
\end{center}
\caption{Asymmetry at $\theta=10^0$ (solid), $\theta=30^0$ (dashed) and
$\theta=150^0$ (dotted) for $q=520 \ {\rm MeV/c}$. The $\omega$ range encompasses both
the QEP and the $\Delta$ domain. The left and right panels respectively refer 
to a vanishing width and to  a finite decay width $\Gamma(s)$.}
\label{fig:asym2new}
\end{figure}

\begin{figure}
\begin{center}
\mbox{\epsfig{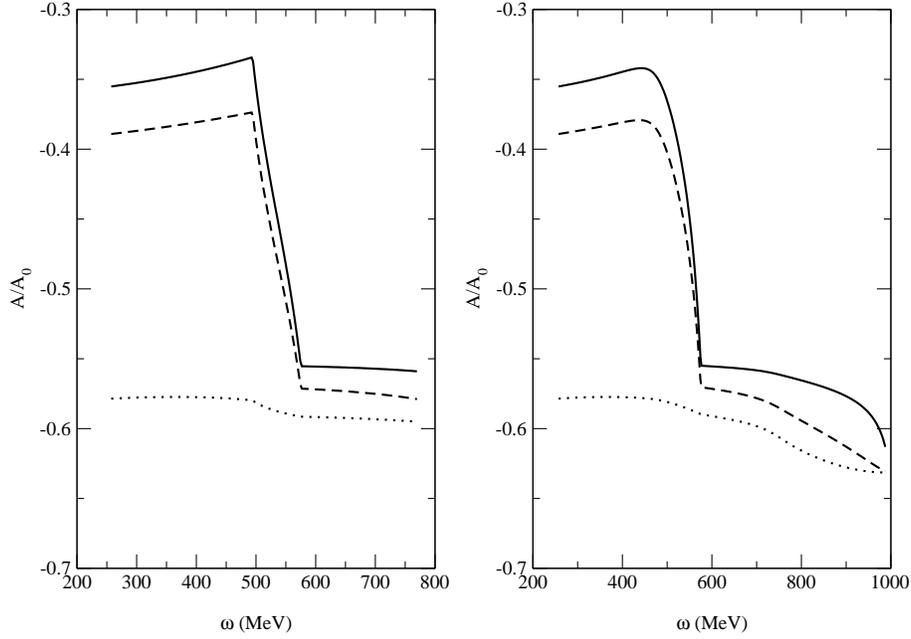}}
\end{center}
\caption{Asymmetry at $\theta=10^0$ (solid), $\theta=30^0$ (dashed) and
$\theta=150^0$ (dotted) for $q=1000 \ {\rm MeV/c}$. The $\omega$ range encompasses both
the QEP and the $\Delta$ domain. The left and right panels respectively refer 
to a vanishing width and to  a finite decay width $\Gamma(s)$.}
\label{fig:asym3new}
\end{figure}

Indeed our results, displayed in the Figs \ref{fig:asym1new}, \ref{fig:asym2new} and
\ref{fig:asym3new}, confirm these expectations. 
There we display the ratio ${\cal A}/{\cal A}_0$ for the same kinematical
conditions considered in calculating the responses in the previous section. 
Indeed it appears that the largest axial contribution occurs at $q = 350 \ 
{\rm MeV/c}$ and $\theta = 10^0$ where it increases the asymmetry by about $10\%$.
On a flat background these effects should be detectable. 
In Figs \ref{fig:asym1new}, \ref{fig:asym2new} and \ref{fig:asym3new} results are 
given both with vanishing width (left panels) and with energy-dependent
width $\Gamma(s)$, taken from \cite{Ama99} (right panels). Only small changes to the
asymmetries are observed.

The most striking result emerging from these figures
relates to the large reduction of the asymmetry in the QEP with respect to 
the $\Delta$ region for small electron scattering angles. 
The interpretation of this result is the following: when $\theta$ is small 
the longitudinal contribution in the QEP makes its most pronounced
contribution to the asymmetry. But this, as we have previously remarked, 
is very small in the RFG framework, because of the isoscalar-isovector
competition. Hence the reduced magnitude of the asymmetry in the QEP
is observed.
In contrast, such a competition does not exist in the $\Delta$ sector, 
which is purely isovector in character.  Hence a large asymmetry occurs, 
independent of the scattering angle. This result appears to us to be
noteworthy: indeed a failure in experimentally detecting it might signal of the
impact of $NN$ and $N\Delta$ correlations.

\begin{figure}
\begin{center}
\mbox{\epsfig{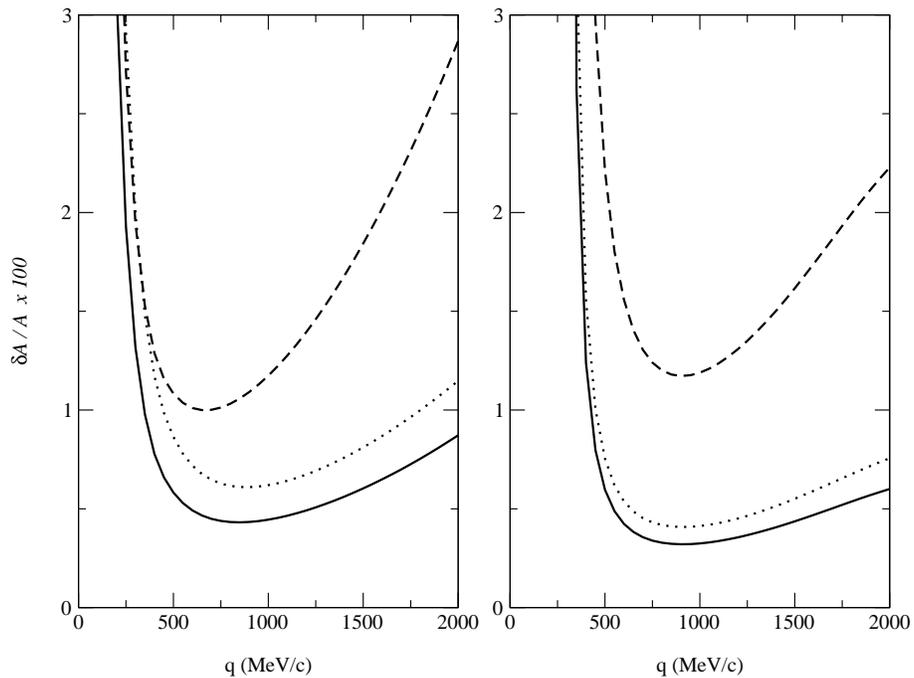}}
\end{center}
\caption{Fractional precision 
$\frac{\delta {\cal A}}{{\cal A}}$ for the $^{12}$C in the QEP 
domain (left) and in the $\Delta$ domain (right). The scattering 
angle takes the three values:
$10^0$ (dotted line), $30^0$ (solid line) and $150^0$ (dashed line). 
We assume the following experimental conditions: $p_e = 1$, 
${\cal L} = 10^{38}$ cm$^{-2}$ s$^{-1}$, $T= 1000$ hr 
and $\Delta\Omega = 250$ msr for $\theta = 30^0$, 150$^0$ and 
$\Delta\Omega = 16$ msr for $\theta = 10^0$. }
\label{fig:prec}
\end{figure}

This being the case, it is important to establish the precision that can be 
reached when measuring the asymmetry in the $\Delta$ region as compared to 
that in the QEP. In fact the two turn out to be close to each other, 
with perhaps the precision in the $\Delta$ sector being even larger, 
as demonstrated in Fig. \ref{fig:prec}, which has been computed for conditions
relevant for CEBAF.

\begin{figure}
\begin{center}
\mbox{\epsfig{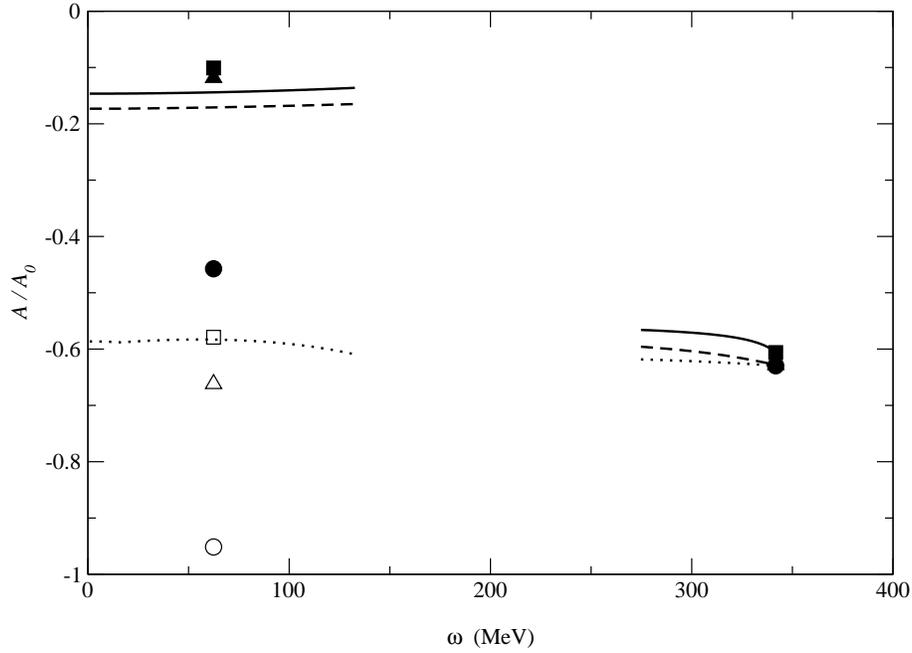}}
\end{center}
\caption{Asymmetry at $q = 350 \ {\rm MeV/c}$. The lines refer to a symmetric
Fermi gas with $k_F = 220 \ {\rm MeV/c}$. The points refer to the asymmetry on a free
nucleon, closed (open) for proton (neutron). 
The solid lines and the squares correspond to $\theta = 10^0$.
The dashed lines and triangles correspond to $\theta = 30^0$. The
dotted lines and circles correspond to $\theta = 150^0$.}
\label{fig:asymq1}
\end{figure}

\begin{figure}
\begin{center}
\mbox{\epsfig{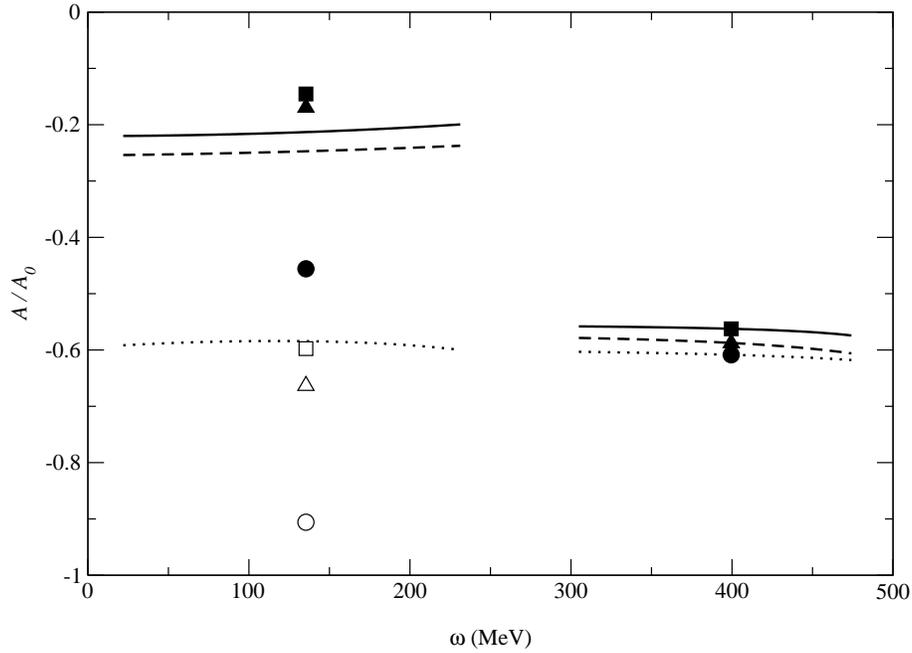}}
\end{center}
\caption{Same as fig. \ref{fig:asymq1} at  $q = 522 \ {\rm MeV/c}$.}
\label{fig:asymq2}
\end{figure}

\begin{figure}
\begin{center}
\mbox{\epsfig{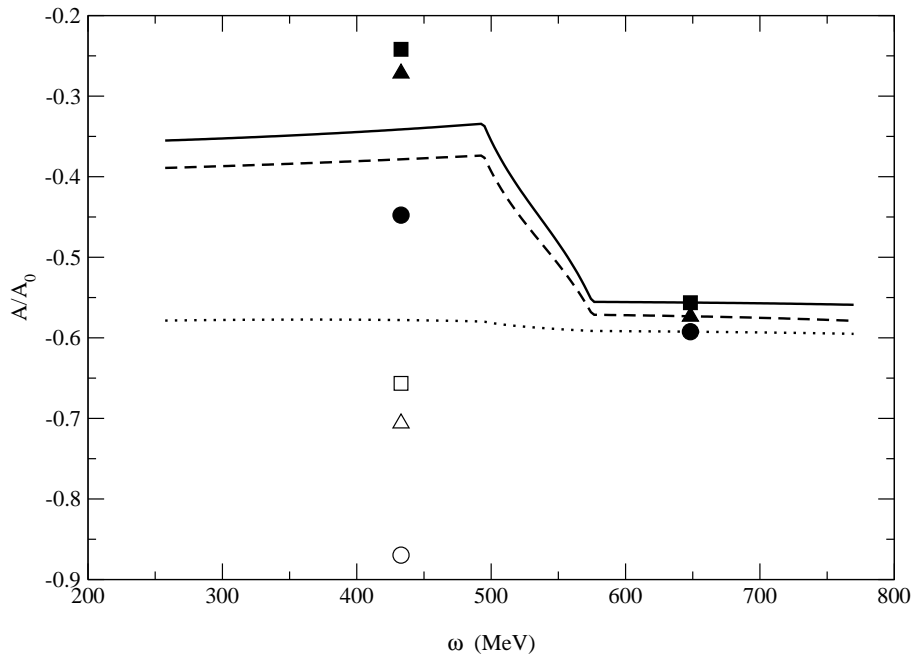}}
\end{center}
\caption{Same as fig. \ref{fig:asymq1} at  $q = 1000 \ {\rm MeV/c}$.}
\label{fig:asymq3}
\end{figure}

In Figs. \ref{fig:asymq1} - \ref{fig:asymq3} we show the RFG asymmetries and for 
comparison the asymmetries found for the proton and neutron under the same 
kinematical conditions.
In the $\Delta$ region all are very similar, as expected, since the 
$N \rightarrow \Delta$ transition is isovector. In contrast, significant 
differences are observed for the quasi-elastic asymmetries. This can
be understood using the following arguments: for scattering in the impulse
approximation we have
\beq
{\cal A}_{nucleus} = \cos^2 \Theta \ {\cal A}_p + \sin^2 \Theta \ {\cal A}_n \ ,
\eeq
where ${\cal A}_{p,n}$ are the individual proton and neutron asymmetries.
Here one has
\beq
\tan^2 \Theta = \frac{N \left[ v_L R^L + v_T R^T\right]_n}{Z 
\left[ v_L R^L + v_T R^T\right]_p} \ .
\eeq
For an $N=Z$ nucleus, as assumed here in obtaining the RFG results, one finds
in the quasi-elastic region that
\beq
\tan^2 \Theta &\approx& \frac{\varepsilon \ G_{E_n}^2 + \tau \
  G_{M_n}^2}{\varepsilon \ 
G_{E_p}^2 + \tau \ G_{M_p}^2} \nonumber \\
&\approx& \frac{\tau \ \mu_n^2}{\varepsilon + \tau \ \mu_p^2} \ ,
\label{eq:theta}
\eeq
and thus for large $\varepsilon$ (small $\theta$) and small $\tau$ we find that 
$\tan^2 \Theta \rightarrow 0$, whereas for large $\varepsilon$ and/or large
$\tau$ we obtain $\tan^2 \Theta \approx \left( \mu_n/\mu_p \right)^2 \approx
4/9$. In (\ref{eq:theta}) $\varepsilon \equiv \left[ 1 + 2 \left( 1 + \tau \right)
\tan^2 \theta/2  \right]^{-1}$.

Using this as a rough guide it is easy to see that the p-to-n weighting of
the single-nucleon asymmetries varies in the required way to explain the
dots and lines in the figures.
As noted in previous work\cite{Don92,Had92}, these differences are very important
when attempting to isolate the various form factor dependencies in the 
quasi-elastic region and now we also have insight into the (different) behaviour
in the $\Delta$ region.

\section{The transverse Coulomb sum-rule}
The deduction of the amount of deformation of the $\Delta$ 
is currently receiving 
a lot of attention experimentally and is of importance for the understanding
of non-perturbative QCD. Accordingly in this Section we study the 
longitudinal response of the $\Delta$, which clearly relates to the above 
issue.
This type of excitation, as already mentioned, gets two contributions: one arises
directly from the Coulomb multipoles (and is small at small $\tau$) and one is
induced by the Fermi motion and relativity. These two elements indeed allow 
the large magnetic excitation of the $\Delta$ to contribute 
in the longitudinal 
channel, the more so the larger is $\tau$.

We start by considering the latter contribution. We do so by {\sl reducing} 
the $R^L$ of the $\Delta$, i.e. by devising a dividing factor such that it  
disentangles, as far as possible, the physics of the single-nucleon from that of
the  many-body problem. This is in complete analogy with the procedure 
adopted in the QEP domain.
We suggest, as a convenient reducing factor, the following expression
\beq
H = \frac{3 A}{4 m_N} \frac{\kappa}{\tau} \ w_2(\tau) \ .
\eeq
Then, by setting $G_C^{(v)}(\tau) \equiv 0$, we obtain for the reduced response
\beq
r^L(\kappa,\lambda) &=& \frac{\xi_F}{\eta_F^3} (1-\psi^2) 
{\cal D}^LL(\kappa,\lambda) \ .
\label{eq:tcsrrl}
\eeq

\begin{figure}
\begin{center}
\mbox{\epsfig{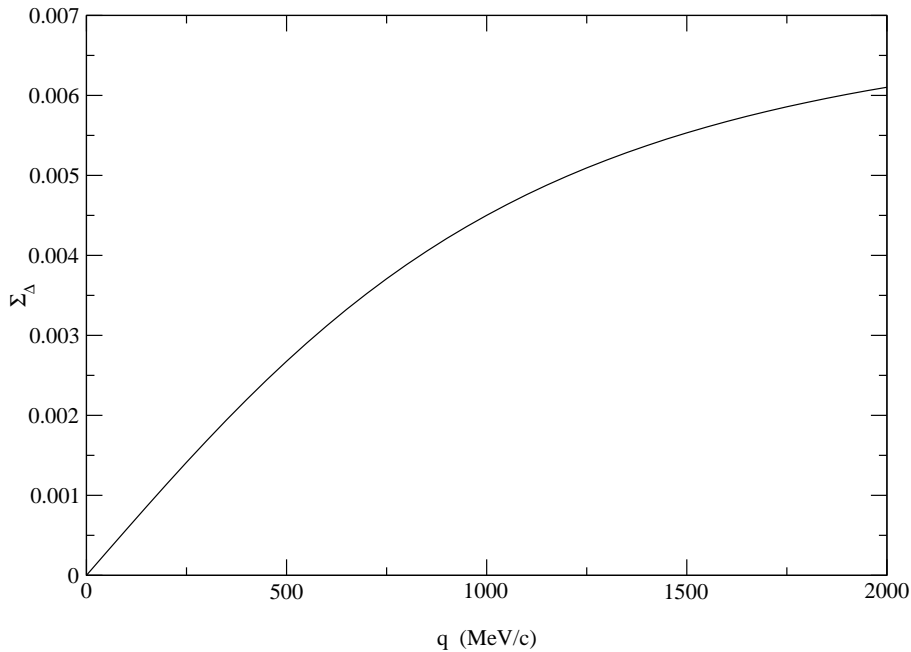}}
\end{center}
\caption{Transverse Coulomb sum rule}
\label{fig:sigmadelta}
\end{figure}

Now a kind of {\sl transverse} Coulomb sum rule $\Sigma_\Delta$ can be worked out
by integrating (\ref{eq:tcsrrl}) over the energy range set by (\ref{eq:regionD}).
To perform this integration is useful to exploit the following integral 
representation for ${\cal D}_L$, namely
\beq
{\cal D}^L(\kappa,\lambda) &=& \frac{1}{\epsilon_F-\tilde{\gamma}_-} 
\int_0^{\eta_F}\ d\eta \ \int_{-1}^{+1} \ d\cos\theta \ \delta \left(\lambda - 
\frac{\epsilon_{\vec{\kappa}+\vec{\eta}}-\epsilon}{2} \right) 
\frac{\kappa \eta^4 \sin^2\theta}{\epsilon \ \epsilon_{\vec{\kappa}+\vec{\eta}}} 
\ ,
\eeq
where we have set
\beq
\epsilon_{\vec{\kappa}+\vec{\eta}} = \sqrt{\mu^2 +\eta^2+4 \kappa^2 + 
4 \kappa \eta \cos\theta} \ .
\eeq
Indeed the $\delta$-function allows us to perform the $\lambda$-integration 
immediately, yielding
\begin{eqnarray}
\Sigma_\Delta &=& \frac{\xi_F}{\eta_F^3}  \int d\lambda (1-\psi^2) 
{\cal D}^L(\kappa,\lambda) \nonumber \\
&=& \frac{\kappa}{\eta_F^3} \int_0^{\eta_F} d\eta \frac{\eta^4}{\sqrt{1+\eta^2}} 
\int_{-1}^{+1} dx \frac{1-x^2}{\sqrt{\zeta^2+\eta^2+4 \kappa^2+4 \kappa \eta x}} \
 \ \ ,
\label{eq:sigmadelta}
\end{eqnarray}
where the integral over the variable $x$ can be analytically expressed in 
terms of elliptic functions. We prefer to perform this task numerically and 
the resulting $\Sigma_\Delta$ is displayed in  Fig. \ref{fig:sigmadelta} for 
$k_F=220 \ {\rm MeV/c}$. In fact, the behaviour of $\Sigma_\Delta(\kappa)$ 
for small and large $\kappa$ can easily be obtained
from (\ref{eq:sigmadelta}) and one gets

\begin{eqnarray}
\Sigma_\Delta &=& \frac{4 \kappa \eta_F^2}{15 \zeta}
\end{eqnarray} 
and
\begin{eqnarray}
\Sigma_\Delta &=& \frac{2 \eta_F^2}{15} \ ,
\end{eqnarray}
respectively.

It thus appears, as expected, that $\Sigma_\Delta$ is a growing
function of the density: should it be an experimentally
accessible quantity, it would allow a determination of the Fermi momentum
$k_F$, which would be interesting to compare with the values obtained via the
width of the quasi-elastic peak or the ground-state density of  nuclei.

\begin{figure}
\begin{center}
\mbox{\epsfig{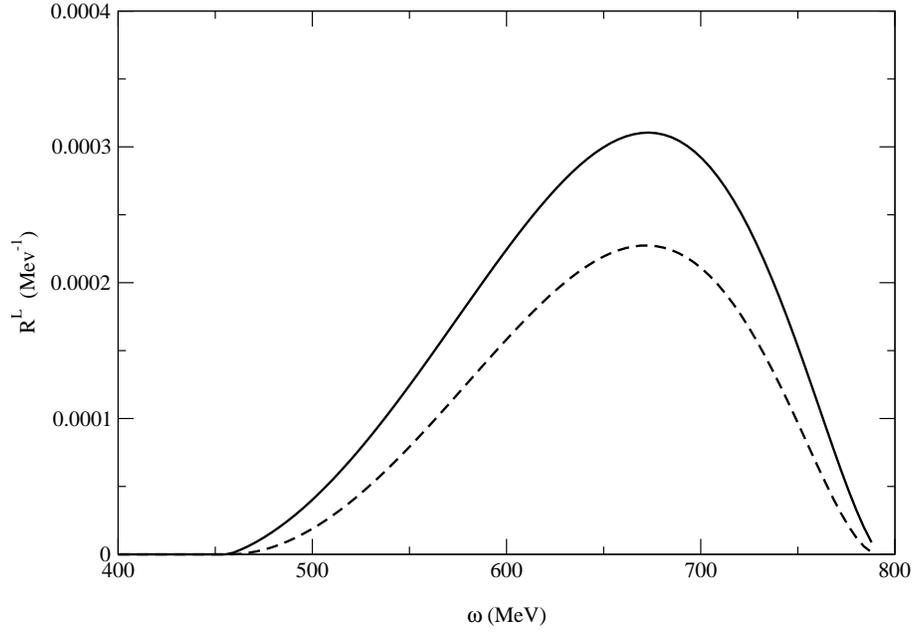}}
\end{center}
\caption{$R^L_{N\Delta}$ with all form factors (solid) and only magnetic (dashed) at
$q = 1000 \ {\rm MeV/c}$.}
\label{fig:rl1}
\end{figure}

\begin{figure}
\begin{center}
\mbox{\epsfig{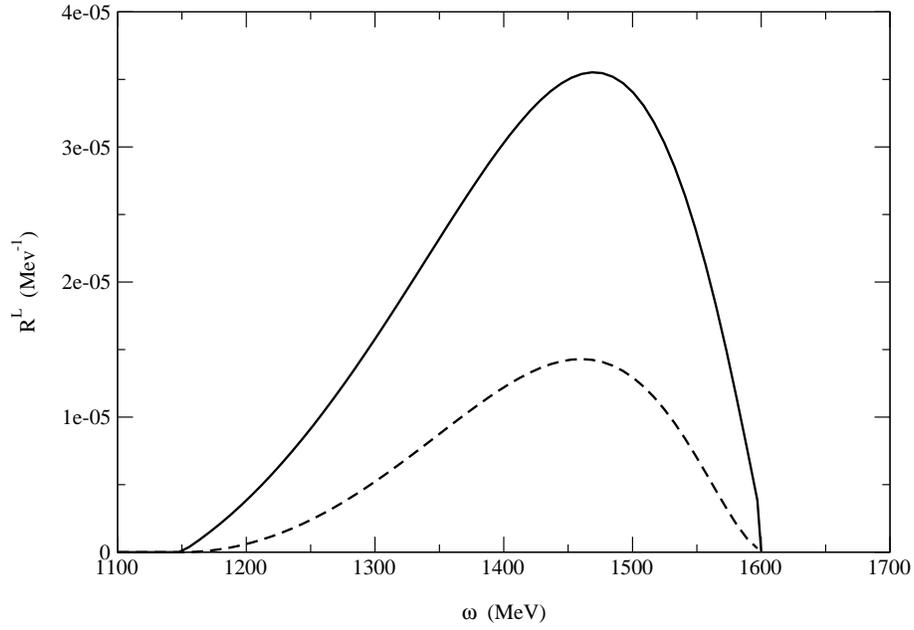}}
\end{center}
\caption{$R^L_{N\Delta}$ with all form factors (solid) and only magnetic (dashed) at
$q = 2000 \ {\rm MeV/c}$.}\
\label{fig:rl2}
\end{figure}

We should now account for the contribution to $R^L$ stemming from
$G_c^{(v)}(\tau)$.  This we do by displaying the full $R^L$ 
(shown together with the separated transverse contribution) 
in Figs. \ref{fig:rl1} and
\ref{fig:rl2} at $q = 1 \ {\rm GeV}$ and $q= 2 \ {\rm GeV}$, respectively, 
and using the Coulomb $N \rightarrow \Delta$ form factor given in 
(\ref{eq:GCV}). We see that while the contribution of the latter remains
moderate at $q = 1 \ {\rm GeV/c}$, it grows strongly as $q$ increases further. 
As already observed in \cite{Ama99}, this casts serious doubts  on the
reliability of the parametrization for the $G_C^{(v)}(\tau)$ adopted in the
present paper. In this way experimental investigation of $\cal A$
could help in elucidating the behaviour of $G_C^{(v)}(\tau)$.

\section{Conclusions}

In this paper we have first studied the responses of nuclei to an external
electromagnetic or weak neutral current field in both the quasi-elastic and 
$\Delta$ domains. We have addressed the formal connections 
between the response 
functions when $m_\Delta$ goes into $m_N$ in the two energy regimes: 
indeed they smoothly evolve from one to the other, but for an appropriate replacement of
the form factors.

On the basis of the computed response functions we have next set up the
left/right asymmetry as measured in the inelastic scattering of longitudinally
polarized electrons off nuclei for momentum transfers up to $1 \ {\rm GeV}$.
We have explored to what extent the $N \rightarrow \Delta$ axial response function stands
out from the otherwise flat energy behaviour characterizing the asymmetry in
the $N \rightarrow \Delta$ sector (i.e., what would occur if the 
axial response vanishes and the
background contributions are negligible).
We have found that for modest momentum transfers and near the light-cone 
an effect exists and should be detectable owing to the high fractional 
precision attainable for measurement of the asymmetry 
in the $\Delta$ domain. This last outcome relates, of course, to the large
cross section for electroexcitation of the $\Delta$.

The most notable feature of the ${\cal A}$  found relates to the
dramatic increase of its magnitude  as one makes a transition from the QEP into the
$N \rightarrow \Delta$ region for small electron scattering angles.
Because of its size this effect should be measurable both at large, say $1 \
{\rm GeV/c}$, and at moderate, say $300-400 \ {\rm MeV/c}$, momentum transfers. 
In the former case nuclear interactions are not likely to disrupt the RFG
predictions too much.
In the latter a modification of the effect could take place, but 
then this might eventually help to shed light on the nature of the $NN$ and $N\Delta$ forces.

Of relevance is also our finding concerning  the proton's and neutron's
asymmetries. It turns out that they differ significantly from each other and
from the RFG results depending upon the specific kinematics. 
Indeed, a comparison between the two as performed here allows one to
identify the kinematical domains where they most differ. Hence it appears
possible that, by measuring the asymmetry on an $N = Z$ nucleus, one
could arrive at information that would help in  disentangling the asymmetry 
on the neutron.

Finally we have explored the longitudinal response function of the $\Delta$, in
particular the interplay between its two contributions involving magnetic and 
Coulomb contributions.
A measurement of the separated $R^L$ in the $\Delta$ domain, while undoubtely
difficult, would greatly improve out knowledge on the elusive nature of the
latter. 

A rich harvest of interesting physics appears indeed to wait to be unraveled
in the $\Delta$ domain.


\section*{Acknowledgments}

This work is supported  in part by DOE grant 333271 (P.A.), 
by the Bruno Rossi INFN-MIT exchange program (R.C. and
A.M.) and by funds provided by the U.S. Department  of Energy under Cooperative 
Research Agreement No. DE-FC02-94ER40818 (T.W.D.).


\end{document}